\RequirePackage{ifpdf}
\ifpdf 
\documentclass[pdftex]{sigma}
\else
\documentclass{sigma}
\fi

\usepackage{dsfont}

\numberwithin{equation}{section}
\numberwithin{theorem}{section}
\numberwithin{definition}{section}
\numberwithin{corollary}{section}

\newcommand{\bem}{\begin{pmatrix}}
\newcommand{\eem}{\end{pmatrix}}

\def\a{\alpha}
\def\b{\beta}
\def\c{\gamma}

\def\e{\epsilon}

\def\g{\gamma}

\def\im{\mathrm{Im}}
\def\inf{\infty}

\def\l{\lambda}
\def\m{\mu}
\def\n{\nu}

\def\p{\pi}

\def\re{\mathrm{Re}}                
\def\r{\rho}                                     
\def\s{\sigma}                                   
\def\t{\tau}

\def\D{\Delta}
\def\F{\Phi}
\def\G{\Gamma}

\def\L{\Lambda}
\def\O{\Omega}

\def\Q{\Theta}

\def\Tr{{\rm Tr}}

\def\cn{{\cal N}}

\def \Z {{\mathbb Z}}

\def \R {{\mathbb R}}

\begin{document}

\allowdisplaybreaks

\renewcommand{\thefootnote}{$\star$}

\renewcommand{\PaperNumber}{068}

\FirstPageHeading

\ShortArticleName{Wall Crossing, Discrete Attractor Flow and Borcherds Algebra}

\ArticleName{Wall Crossing, Discrete Attractor Flow\\ and Borcherds Algebra\footnote{This paper is a
contribution to the Special Issue on Kac--Moody Algebras and Applications. The
full collection is available at
\href{http://www.emis.de/journals/SIGMA/Kac-Moody_algebras.html}{http://www.emis.de/journals/SIGMA/Kac-Moody{\_}algebras.html}}}

\Author{Miranda C.N. CHENG~$^\dag$  and Erik P. VERLINDE~$^\ddag$}

\AuthorNameForHeading{M.C.N. Cheng and E.P. Verlinde}

\Address{$^\dag$~Jef\/ferson Physical Laboratory, Harvard University, Cambridge, MA 02128, USA}
\EmailD{\href{mailto:mcheng@physics.harvard.edu}{mcheng@physics.harvard.edu}}

\Address{$^\ddag$~Institute for Theoretical Physics, University of Amsterdam,\\
$\phantom{^\ddag}$~Valckenierstraat 65, 1018 XE, Amsterdam, the Netherlands}
\EmailD{\href{mailto:e.p.verlinde@uva.nl}{e.p.verlinde@uva.nl}}

\ArticleDates{Received July 01, 2008, in f\/inal form September
23, 2008; Published online October 07, 2008}

\Abstract{The appearance of a generalized (or Borcherds--) Kac--Moody algebra in the spectrum of BPS dyons in $\cn=4$, $d=4$ string theory is elucidated. From the low-energy supergravity analysis, we identify its root lattice as the lattice of the $T$-duality invariants of the dyonic charges, the symmetry group of the root system as the extended $S$-duality group $PGL(2,\Z)$ of the theory, and the
walls of Weyl chambers as the walls of marginal stability for the relevant two-centered solutions. This leads to an interpretation for the Weyl group as the group of wall-crossing, or the group of discrete attractor f\/lows. Furthermore we propose an equivalence between
a ``second-quantized multiplicity'' of a charge- and moduli-dependent highest weight vector and the dyon degeneracy, and show that the wall-crossing formula following from our proposal agrees with the wall-crossing formula obtained from the supergravity analysis. This can be thought of as providing a microscopic derivation of the wall-crossing formula of this  theory.}


\Classification{81R10; 17B67}

\renewcommand{\thefootnote}{\arabic{footnote}}
\setcounter{footnote}{0}

\section{Introduction}

\looseness=1
The study of BPS states in string theory and their relation to the micro-state counting of extremal black holes has been an active and fruitful f\/ield of research for more than a decade. Often these studies were aimed at making a comparison between the degeneracies of BPS states with the macroscopic Bekenstein--Hawking entropy. For that specif\/ic purpose it is suf\/f\/icient to know the counting only in its asymptotic form, but eventually one would like to arrive at a~more complete and detailed microscopic theory of  black hole states and extract more physical information than just the asymptotic growth of the counting. In this respect the $1/4$-BPS black holes in $4d$ string theories with $\cn =4$ space-time supersymmetry are special, since they have been the f\/irst and long been the only case  for which an exact microscopic counting formula was known. So far no complete formula exists for  theories with fewer than four supersymmetries.

Related to the counting of BPS states, various ideas have been put forward throughout the years. There was the pioneering idea that a Borcherds--Kac--Moody algebra should be related to the BPS sector of a supersymmetric string theory\footnote{Although the analysis in these papers concentrates on the perturbative BPS states, while the BPS states considered in the present paper are non-perturbative $1/4$-BPS states.} \cite{Harvey:1995fq,Harvey:1996gc}, and the recent momentum in the study of the moduli-dependence of the BPS spectrum \cite{Denef:2007vg,Diaconescu:2007bf,Kontsevich_Soibelman}. The $\cn =4$, $d=4$ theory has been shown to be a suitable playground for these ideas. Firstly, it was known from the start that the dyon-counting formula proposed in  \cite{Dijkgraaf:1996it} is related to the denominator formula of a Borcherds--Kac--Moody superalgebra \cite{Kawai:1995hy,GN1}. Secondly, thanks to the simplif\/ication brought by higher supersymmetries, the moduli-dependence of the BPS states in the $\cn =4$, $d=4$ theories has been extensively studied in the last two years \cite{Sen:2007vb,Dabholkar:2007vk,Cheng:2007ch,Sen:2007nz,Dabholkar:2008zy,Sen:2007pg}. In this paper we will see how the two aspects of the $\cn=4$, $d=4$ dyon spectrum are intimately related to each other.

For the $\cn =4$, $d=4$ string theory obtained by toroidally compactifying the heterotic string, or equivalently compactifying type II string on $K3\times T^2$, an exact dyon-counting formula
\begin{gather}
\label{DVV0}
(-1)^{P\cdot Q+1} D(P,Q) =\oint d\O \,\frac{e^{\p i (P^2 \r + Q^2 \s - 2 P\cdot Q \n)} }{\F(\O)},\qquad \O = \bem\s & \n \\ \n & \r \eem
\end{gather}
was f\/irst proposed in \cite{Dijkgraaf:1996it} and later improved in \cite{Shih:2005uc}, where $P$ and $Q$ are vectors in the unique even unimodular (or self-dual) 28-dimensional Lorentzian charge lattice $\G^{6,22}$, and
$P^2$, $Q^2$ and~$P\cdot Q$ are the three $T$-duality invariants obtained by combining the charge vectors using the natural $O(6,22;\Z)$-invariant bilinear form, and
the generating function $\F(\O)^{-1}$ has an inf\/inite product representation due to its relation to the Weyl--Kac--Borcherds denominator of a certain generalized (or Borcherds--) Kac--Moody algebra.
More recently, the above formula has been derived using the known results of the BPS degeneracies of the D1-D5 system \cite{Shih:2005uc,David:2006yn}.
The counting formula contains more information than its asymptotic agreement with
the known  Bekenstein--Hawking entropy \cite{Cvetic:1995bj}
\begin{gather}\label{entropy}
S= \p \sqrt{P^2 Q^2 -(P\cdot Q)^2}.
\end{gather}
In particular, it has been demonstrated \cite{Sen:2007vb,Cheng:2007ch} that the moduli dependence of the spectrum as predicted by the macroscopic supergravity analysis can also be correctly incorporated in the above counting formula.
Nevertheless, despite all the progress, no explanation has been given for the mysterious appearance of the Borcherds--Kac--Moody algebra,
neither from a microscopic nor from a macroscopic point of view.
Understanding the origin of this algebra from either direction can be an important step
towards developing a complete microscopic theory for these extremal dyonic black holes.

The aim of this paper is to elucidate the role of this algebra in the description of the $\cn=4$ BPS dyons. First, we will show that just from a purely macroscopic perspective, i.e.\ from the low-energy supergravity theory,  one can understand the appearance of its root lattice, its hyperbolic Weyl group, and discover the intricate stucture of all its Weyl chambers. Such a hyperbolic Weyl group has also appeared in gravitational physics in other contexts, see \cite{Damour:2002et,Henneaux:2007ej} and references therein.
Second, using the known denominator formula we will propose an equivalence between  the
dyon degeneracy and a ``second-quantized multiplicity''
of a  charge- and moduli-dependent highest weight. From this perspective a change of moduli can be seen as a change of the ``vacuum state'', or equivalently a change of representation of the algebra.

In more details, the $T$-duality invariants of the dyonic charges naturally form  a three-dimensional Lorentzian lattice, in which future-pointing light-like vectors correspond to the $1/2$-BPS while the time-like ones represent the  $1/4$-BPS states.
This lattice is exactly the root lattice of the well-known generalized Kac--Moody algebra constructed by Gritsenko and Nikulin~\cite{GN1}.
Moreover, the $PSL(2,\Z)$ $S$-duality group of the theory can be extended to $PGL(2,\Z)$ by inclu\-ding the spatial parity transformation. This group turns out to be exactly the symmetry group of the root system of the above Weyl group.

Furthermore,  by studying the stability condition for the relevant two-centered solutions in the low-energy supergravity theory, we conclude that these solutions are labelled by the positive real roots of the algebra, and the walls of marginal stability can be identif\/ied with the walls of Weyl chambers. When the background moduli is changed to move along the attractor f\/low of a black hole to the attractor moduli, generically various walls of marginal stability will be crossed in a~certain order and the corresponding two-centered solutions will disappear from the spectrum. Therefore we can describe the discrete moduli dependence of the spectrum as a sequence of Weyl ref\/lections, and identify the Weyl group of the algebra as a ``group of discrete attractor f\/lows''. See \cite{Ferrara:1995ih,Mohaupt:2000mj} for the explanation of the attractor mechanism in a supersymmetric black hole background.

\looseness=1
This supergravity analysis is directly mirrored in the microscopic counting formula, which expresses the BPS degeneracies in terms of the denominator of the Borcherds--Kac--Moody al\-geb\-ra.
First of all, it was shown in \cite{Cheng:2007ch} that the ef\/fect of wall-crossing can be incorporated in the counting formula through a choice of contour that depends on the moduli. By comparing the dyon-counting formula equipped with this contour prescription with the character formula for the Verma module of the algebra, we show how the dyon degeneracy is given by the number of ways in which a corresponding highest weight can be written as a sum of positive roots. Under this identif\/ication, the moduli-dependent prescription for integration contours is equivalent to a~moduli-dependent prescription for the highest weight of the relevant Verma module.

Furthermore, the hierarchy (i.e.\ up- vs. down-stream) of the attractor f\/low induces a hierarchy of the corresponding Verma modules and we will see how the representation becomes ``smaller'' when the background moduli change along an attractor f\/low moving towards the attractor point.
From this consideration, and
using the properties of the multiplicities of the roots of the algebra, one can show that the dif\/ference coincides with the degeneracy of the corresponding two-centered solution as predicted using supergravity. This can be thought of as providing a~microscopic derivation of the wall-crossing formula.

In this paper we will assume that both the magnetic and the electric charge vectors are primitive in the lattice $\G^{6,22}$, in particular that they satisfy the so-called ``co-prime condition''
\begin{gather}\label{co_prime_condition}
\text{g.c.d}(P^a Q^b - P^b Q^a) = 1,\qquad a,b = 1,\dots,28,
\end{gather}
which ensures that the degeneracies are completely determined by the set of three $T$-duality invariants ($P^2/2, Q^2/2, P\cdot Q$). When this is not true there are extra possible multi-centered solutions which should be considered.
Please see \cite{Maldacena:1999bp,Banerjee:2008ri,Dabholkar:2008zy} for the relevant discussions.

This paper is organized as follows.
In Section~\ref{sec2} we study the Lorentzian  lattice of  the vectors of $T$-duality invariants  and see
how a basis of this lattice def\/ines a hyperbolic ref\/lection group, which coincides with the Weyl group of the Borcherds--Kac--Moody algebra known to be related to the BPS states of this theory. We then
show that the extended $S$-duality group of the theory coincides with the symmetry group of the root system of this Weyl group. In Section~\ref{sec3} we f\/irst review the supergravity analysis of the walls of marginal stability of the relevant two-centered solutions, and establish their one-to-one correspondence with the positive roots of the Weyl group def\/ined in Section~\ref{sec2}.
Using the results in the previous two sections, in Section~\ref{sec4} we will interpret the Weyl group as the group of discrete attractor f\/lows, and discuss the implication of the presence of such a group structure underlying the attractor f\/low for the phenomenon of wall-crossing in this theory.
In Section~\ref{sec4.3} we give a simple arithmetic description of this attractor group and show how the attractor f\/low can be identif\/ied with a process of coarse-graining the rational numbers. In Section~\ref{sec5} we move on to discuss the role played by the Borcherds--Kac--Moody algebra under consideration in the BPS spectrum of the theory. After some background mathematical knowledge is introduced in Section~\ref{sec5.1}, in Section~\ref{sec5.2} we see how a jump across the wall is related to a change of representation of the algebra. Using this prescription,
in Section~\ref{sec5.3} we present a microscopic derivation of the wall-crossing formula using the knowledge of the root multiplicity of the algebra.

\section{Dyons and the Weyl group}\label{sec2}

In this section we will show that the vector of the three quadratic invariants $P^2$,  $Q^2$, $Q \cdot  P$ lives naturally in a Lorentzian lattice of signature $(+,+,-)$, and a basis of this lattice generates a~group of ref\/lection which, as we will see later, coincides with the Weyl group of the Borcherds--Kac--Moody algebra. Afterwards we will brief\/ly discuss the physical relevance of this group while leaving the details for later sections.

\subsection{The root lattice}\label{sec2.1}
The fact that the invariants  naturally form a Lorentzian lattice can  be seen from the entropy formula (\ref{entropy}) for example, as well as from the fact that $S$-duality acts on them as the discrete Lorentz group $SO^+(2,1;\Z)\sim PSL(2,\Z)$, where the ``+'' denotes the time-orientation preserving component of the Lorentz group. To make this Lorentz action more explicit,  it is convenient to collect the three invariants $P^2$, $Q^2$ and $P\cdot Q$
 into a symmetric matrix. Indeed,  the (2+1)-dimenional Minkowski space $\R^{2,1}$ can be naturally identif\/ied with the space of $2 \times 2$ symmetric matrix with real entries
\begin{gather*}
 \left\{X\Big\lvert \;X= \bem x_{11} & x_{12} \\ x_{21} & x_{22} \eem,\; X=X^T,\; x_{ab} \in \R\right\}
\end{gather*}
equipped with metric\footnote{The ``$-2$'' factor here and in various places later is due to the fact that we choose the normalization of the metric to be consistent with the familiar convention for Kac--Moody algebras in which the length squared of a~real simple root is 2 for cases that all the real simple roots have the same length.}
\begin{gather*}
(X,X)= -2 \det X .
\end{gather*}
For later notational convenience we will also def\/ine $ |X|^2= -(X,X)/2 =\det X  $.

The corresponding inner product is
\[
(X,Y) =  - \e^{ac}\e^{bd} x_{ab} y_{cd} = -\det Y \,\Tr\big(XY^{-1}\big),
\]
where $\e$ is the usual epsilon symbol $\e^{12} = -\e^{21} =1$.
Furthermore, we will choose the direction of time in this space to be such that the future light-cone is given by
\begin{gather*}
V^+ = \left\{X\Big\lvert \;X= \bem x_{11} & x_{12} \\ x_{21} & x_{22} \eem,\; X=X^T, \;x_{ab} \in \R, \;\Tr X > 0 , \;\det X > 0\right\}.
\end{gather*}

One can immediately see that this is indeed a vector space of signature (2,1), in which the diagonal entries of the $2\times 2$ matrix play the role of the two light-cone directions. As mentioned earlier, an element of the $S$-duality group $PSL(2,\Z)$ acts as a Lorentz transformation on this space:
for any real matrix $\g$ with determinant one, one can check that the following action
\begin{gather}\label{action_PGL2Z}
X \to \g(X) := \g X \g^T
\end{gather}
is indeed a Lorentz transformation satisfying $ |X|^2 =|\g(X)|^2$.

A natural vector to consider in this vector space is the ``charge vector'' made up of the three quadratic invariants appearing in the dyon-counting formula (\ref{DVV0})
\begin{gather*}
\L_{P,Q} = \bem P\cdot P & P\cdot Q\\ P\cdot Q & Q\cdot Q \eem.
\end{gather*}
Using this metric, the entropy of a dyonic black hole (\ref{entropy}) becomes nothing but given by the length of the charge vector $\L_{P,Q}$ as
\begin{gather*}
S(P,Q) =\p|\L_{P,Q}|.
\end{gather*}

Next we would like to consider a basis for the charge vectors $\L_{P,Q}$. From the fact that $P^2$, $Q^2$ are both even, it is easy to check that for any dyonic charge which permits a black hole solution, namely for all $(P, Q)$ with $S(P,Q)>0$, the charge vector $\L_{P,Q}$ is an integral linear combination with non-negative coef\/f\/icients of the following basis vectors
\begin{gather}\label{3_simple_roots}
\a_1 =\left(\!\!\begin{array}{rrr} 0 &-1 \\ -1& 0 \end{array}\!\!\right),\qquad
\a_2 =\bem 2 &1 \\ 1& 0 \eem,\qquad
\a_3 =\bem 0 &1 \\ 1& 2 \eem.
\end{gather}
In other words, for all black hole dyonic charges we have
\begin{gather*}
\L_{P,Q} \in \{\Z_+\a_1+\Z_+\a_2+\Z_+\a_3 \} .
\end{gather*}

The matrix of the inner products of the above basis is
\begin{gather*}
(\a_i,\a_j) = \left(\!\! \begin{array}{rrr}2&-2&-2\\ -2&2&-2\\ -2&-2&2\end{array}\!\!\right).
\end{gather*}

Later we will see that this is the real part, namely the part with positive diagonal entries, of the generalized Cartan matrix of the relevant Borcherds--Kac--Moody algebra.

\subsection{The Weyl group}\label{sec2.2}
We can now def\/ine in the space $\R^{2,1}$ the group $W$ generated by the ref\/lections with respect to the spacelike vectors $\a_{1,2,3}$:
\begin{gather}\label{generator_weyl}
s_i : \ X \to X - 2\frac{(X,\a_i)}{(\a_i,\a_i)}\a_i,\qquad i=1,2,3.
\end{gather}
Note that there is no sum over $i$ in the above equation.

\begin{figure}[t]
\centering
\includegraphics[width=3cm]{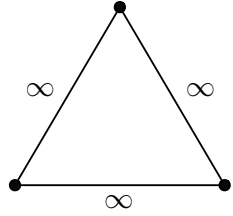} 
\caption{The Coxeter graph of the hyperbolic ref\/lection group generated by (\ref{generator_weyl}). See (\ref{def_coxeter}) for the def\/inition of the Coxeter graph.}\label{Coxeter_graph}
\end{figure}

This group turns out to be a hyperbolic Coxeter group with the Coxeter graph shown in Fig.~\ref{Coxeter_graph}. The ``$\inf$'' in the graph denotes the fact that the composition of any two distinct basic ref\/lections is a parabolic group element, namely $(s_is_j)^\inf = 1$ for all $i\neq j$.
 The def\/inition and basic properties of Coxeter groups can be found in the Appendix~\ref{Appendix: Properties of Coxeter Groups}. We will from now on refer to this group as the Weyl group, anticipating the role it plays in the Borcherds--Kac--Moody superalgebra discussed in the following sections. In particular, we will denote as $\D^{\rm re}_+ $ the set of all positive roots of the Weyl group (\ref{root1})
\begin{gather}\label{real_roots}
\D^{\rm re}_+ =  \{\a = w (\a_i),w\in W, i =1,2,3 \} \cap \{\Z_+\a_1+\Z_+\a_2+\Z_+\a_3 \} ,
\end{gather}
as it will turn out to be the set of all {\it real} (i.e.\ spacelike) positive roots of the Borcherds--Kac--Moody algebra under consideration.

The physical relevance of this group can be intuitively understood in the following way. We know that the $S$-duality group $PSL(2,\Z)$ is a symmetry group of the theory. We can further extend this symmetry group with the spacetime parity reversal transformation
\begin{gather}\label{parity_reversal}
\l \to - \bar{\l},\qquad \bem P\\Q\eem \to \bem P\\-Q\eem
\end{gather}
and thereby extend the group $PSL(2,\Z)$ to $PGL(2,\Z)$, where ``$P$'' denotes the fact that, as  $2\times 2$ matrices, the extra identif\/ication $\g \sim -\g$ is imposed. Notice that the requirement that the inverse of an element $\g \in PGL(2,\Z)$ is also an element implies $\det\g=\pm 1$.

Explicitly, this group acts on the charges and the (heterotic) axion-dilaton as
\begin{gather}\nonumber
\bem P\\Q\eem \to \bem a&b\\c&d \eem \bem P\\Q\eem,\\
\label{extended_S-dual}
\l \to  \left\lbrace\begin{array}{ll}\displaystyle \frac{a\l+b}{c\l+d} \quad & \text{when} \ ad-bc=1,\\[4mm]  \displaystyle \frac{a\bar{\l}+b}{c\bar{\l}+d}\quad &  \text{when}\  ad-bc=-1. \end{array} \right.
\end{gather}

From the point of view of the Lorentzian space $\R^{2,1}$, the above parity-reversal element,
when acting as (\ref{action_PGL2Z}),
 augments the restricted Lorentz group with the spatial ref\/lection.

As we will prove now, this extended $S$-duality group is also the symmetry group of the root system $\D^{\rm re}$ of the Weyl group def\/ined above, which can be written as the semi-direct product of the Weyl group $W$ and the automorphism group of its fundamental domain (the fundamental Weyl chamber), def\/ined as
\begin{gather*}
{\cal W} :=  \big\{ x \in \R^{2,1}, \   (x,\a_i) < 0,\ i=1,2,3\big\}.
\end{gather*}

\begin{figure}[t]
\centering
\includegraphics[width=10cm]{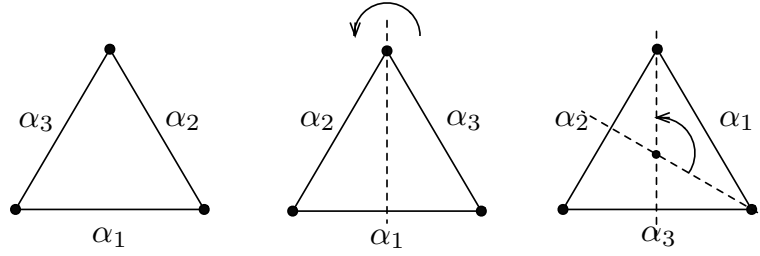} 
\caption{The dihedral group $D_3$, which is the symmetry group of an equilateral triangle, or the outer automorphism group of the real roots of the Borcherds--Kac--Moody algebra (the group of symmetry mod the Weyl group), is generated by an order two element corresponding to a ref\/lection and an order three element corresponding to the $120^\circ$ rotation.}\label{dihedral_grp}
\end{figure}

The automorphism group of the fundamental Weyl chamber is in this case the dihedral group $D_3$ that maps the equilateral triangle, bounded by the planes of orthogonality to $\{\a_1,\a_2,\a_3\}$, to itself. Explicitly, the $D_3$ is the group with six elements generated by the following two generators: the order two element corresponding to the ref\/lection
\begin{gather} \label{dihedral_action_1}
\a_1 \to \a_1 ,\qquad \a_2 \leftrightarrow \a_3
\end{gather}
and order three element corresponding to the $120^\circ$ rotation
\begin{gather}\label{dihedral_action_2}
\a_1 \to \a_2 ,\qquad \a_2 \to \a_3,\qquad \a_3 \to \a_1
\end{gather}
of the triangle. See Fig.~\ref{dihedral_grp}.

Recall that the usual $S$-duality group $PSL(2,\Z)$ is generated by the two elements $S$ and $T$, satisfying the relation $S^2=(ST)^3 =1$. In terms of $2\times 2$ matrices, they are given by
\[
S =\bem0&-1\\ 1&0\eem,\quad T=\bem1 & 1\\ 0&1\eem.
\]
The extended $S$-duality group $PGL(2,\Z)$ is then generated by the above two generators, together with the other one corresponding to the parity reversal transformation (\ref{parity_reversal})
\[
R = \bem 1&0\\0&-1 \eem.
\]

On the other hand, in terms of these matrices and the $PGL(2,\Z)$ action (\ref{action_PGL2Z}) on the vectors in the Lorentzian space, the generator of the Weyl group $W$ corresponding to the ref\/lection with respect to the simple root $\a_1$, is given by
\begin{gather*}
s_1: X \to R(X).
\end{gather*}
For the dihedral group $D_3$,
the ref\/lection generator exchanging $\a_2$ and $\a_3$ is given by
\begin{gather*}
X\to RS(X)
\end{gather*}
and the order three $120^\circ$ rotation generator is given by
\begin{gather*}
X\to TS(X).
\end{gather*}

From the expression for these three generators one can deduce the rest of the elements of $W$. For example, the ref\/lections $s_2$, $s_3$ with respect to the other two simple roots $\a_2$, $\a_3$ are given by $R$ conjugated by the appropriate power (1 and 2 respectively) of the rotation generator $TS$.

In particular, we have shown that the extended $S$-duality group can be written as
\begin{gather*}
 PGL(2,\Z) \cong O^+(2,1;\Z)\cong W \rtimes D_3.
\end{gather*}
This means that the Weyl group is a normal subgroup of the group $ PGL(2,\Z)$, namely that the conjugation of a Weyl group element with any element of $PGL(2,\Z)$ is again a Weyl group element.

In particular, the extended $S$-duality group coincides with the symmetry group of the root system of the Weyl group, or the real root system of the Borcherds--Kac--Moody algebra which will be introduced in Section~\ref{sec5}. This relation with the physical symmetry group suggests a~role of the Weyl group in the physical theory. Indeed, its relation with
 the physical phenomenon of crossing the walls of marginal stability will be clarif\/ied in the next section.

\section{Walls of marginal stability}\label{sec3}

In this section we will show that the walls of the marginal stability of the relevant two-centered solutions of the present theory can be identif\/ied with the walls of Weyl chambers of the Weyl group introduced in the previous section. In particular, the dif\/ferent two-centered solutions with two $1/2$-BPS centers are labelled by the positive roots of the Weyl group, which are the positive real roots of the Borcherds--Kac--Moody algebra which will be introduced in Section~\ref{sec5}.

\subsection{Determining the walls}\label{sec3.1}

To introduce  the problem of interest and to set up the notation, we begin with a short review of  $\cn=4$ two-centered solutions and their associated walls of marginal stability.
We consider the ${\cal N}=4$, $d=4$  supergravity theory corresponding to heterotic string theory compactif\/ied on $T^6$, or equivalently type II on $K3\times T^2$.
The moduli space of the supergravity theory is
\begin{gather*}
{\cal M} = {\cal M}_0 / G,
\end{gather*}
where
\begin{gather*}
{\cal M}_0 = \frac{O(6,22,\R)}{O(6,\R)\times O(22,\R)} \times \frac{SL(2)}{U(1)}
\end{gather*}
is its covering space, and
\begin{gather*}
G= O(6,22,\Z) \times SL(2,\Z)
\end{gather*}
is the $U$-duality group consisting of $T$-duality and $S$-duality transformations.
The $T$-duality group only acts on f\/irst factor in ${\cal M}_0$, which in the heterotic picture represents the Narain moduli.
The second factor, which is the same as the upper-half (or Lobachevski) plane, describes the heterotic axion-dilaton f\/ield $\l=\l_1+ i \l_2$.

We are interested in the dyonic BPS states respecting one-fourth of the sixteen space-time supersymmetries.  There are also $1/2$-BPS solutions, which have no horizon at the leading order. Indeed, all $1/2$-BPS states can be mapped to purely electric states using $S$-duality transformation, which are perturbative heterotic string states with partition function
\begin{gather}\label{half_BPS}
d(P)= \oint d\s \, \frac{e^{-\p i P^2 \s}}{\eta^{24}(\s)},
\end{gather}
where $\eta(\s)$ is the Dedekind eta-function
\[
\eta(\s) = q^{1/24}\prod_{n=1} ^{\inf} (1-q^n) ,\qquad q= e^{2\p i \s}.
\]
See  \cite{Dabholkar:1989jt,Dabholkar:2004dq} for the microscopic counting of the $1/2$-BPS states and the corresponding supergravity solutions.

Apart from the usual static black hole solution, there are also $1/4$-BPS multi-centered solutions. For our purpose of studying the moduli dependence of the spectrum, it is suf\/f\/icient to consider the following special class of them \cite{Sen:2007nz}.  They are $1/4$-BPS solutions consisting of two centers separated by a coordinate distance that is f\/ixed by the charges and the moduli, with each center carrying the charge of a $1/2$-BPS object.
Specif\/ically, this means that for each center the electric and magnetic charge vectors are parallel to each other in the charge lattice~$\Gamma^{6,22}$ and that each center is horizonless at the leading order and in particular does not carry a macroscopic entropy.

These two-centered solutions exist only in a certain region of the moduli space $\cal M$, which depends on the charges $P$ and $Q$. The boundary of these regions are called ``the walls of marginal stability'' and represent the location in moduli space where the $1/4$-BPS  states can decay into inf\/inity and disappear from the spectrum.
This is only possible when the BPS mass of the $1/4$-BPS  solution equals the sum of the individual masses of each $1/2$-BPS object.

Let us illustrate this for the two-centered solution corresponding to the simplest split of charges into the electric and the magnetic part
\begin{gather*}
\bem P \\ Q \eem =\bem P \\ 0 \eem+\bem 0 \\ Q \eem.
\end{gather*}

By quantizing the two-particle system \cite{Denef:2002ru} one f\/inds that the conserved angular momentum is given by
\[
2J+1 = |P\cdot Q| ,
\]
and thereby concludes that the graded degeneracies of the above two-centered solution are given~by
\begin{gather}\label{degeneracy_simple_split}
 (-1)^{P\cdot Q+1}\, |P\cdot Q|\,d(P)\,d(Q) ,
\end{gather}
where $d(P)$, $d(Q)$ are the degeneracies of the two $1/2$-BPS centers given by (\ref{half_BPS}).

From the above discussion, we know that marginal stability must happen on the co-di\-men\-sio\-nal one subspace of the moduli space where  the BPS mass of the dyon with charge $(P,Q)$ saturates  the triangle inequality
\[
M_{P,Q} = M_{P,0} + M_{0,Q}.
\]
We would like to know where in  the moduli space this happens.
The BPS mass for a dyon depends on the charges, the axion-dilaton  moduli and also the Narain moduli.  The latter can be
represented as an $6\times 28$ matrix $\m$, which projects the 28 component charge vectors $P$ and $Q$ onto their six-dimensional left-moving part. We write $P_L= \m P$ and and similarly for the $Q$'s.

After some straightforward algebra one f\/inds that the condition of marginal stability is given by \cite{Sen:2007vb}
\begin{gather}
\label{simplewall}
{ \lambda_1\over \lambda_2} + {P_L \cdot  Q_L\over |P_L \wedge  Q_L|} =0 ,
\end{gather}
where we have def\/ined the wedge product such that $ |P_L \wedge  Q_L| = \sqrt{Q_L^2 P_L^2 - (Q_L\cdot P_L)^2}$.

Note that the Narain moduli only enters through the inner products of $P_L$ and $Q_L$, and therefore this wall is tangent to any variation of these moduli that leaves these inner products invariant. In fact, this observation is also true for the walls of marginal stability associated with  more general two centered solutions.

The next step will be to consider the other ways in which a dyon can split into two $1/2$-BPS particles, and determine the
corresponding walls of marginal stability.
It is easy to check that every element of $PGL(2,\Z)$ gives the following split of charges
\begin{gather}\label{split_1}
\bem P \\ Q \eem = \bem P_1 \\ Q_1 \eem +\bem P_2 \\ Q_2 \eem
=  \pm\left\lbrack(-cP+dQ)    \bem b \\ a\eem+ (aP-bQ)    \bem d \\ c\eem  \right\rbrack  .
\end{gather}
where the plus (minus) sign holds depending on whether   $ad - bc  =   1$ or $-1$.
Furthermore, from the quantisation condition of the charges and the fact that a $1/2$-BPS charges must have the magnetic and electric charges parallel to each other, one can also show that the converse is also true \cite{Sen:2007vb}. Namely,  for every possible $1/2$-BPS split one can always f\/ind a (not unique) $ PGL(2,\Z)$ element
such that the charges can be  written in the above form.

Using similar analysis, we can determine the wall of marginal stability of the two-centered solution with the split of charges corresponding to a generic element in $ PGL(2,\Z)$
to be \cite{Sen:2007vb}
\begin{gather}
\label{wallgamma}
ac \left(\frac{|\l|^2}{\l_2} + \frac{P_L\cdot P_L}{|P_L\wedge Q_L|}\right)
- (ad+bc)\, \left(\frac{\l_1}{\l_2} + \frac{P_L\cdot Q_L}{|P_L\wedge Q_L|}\right) + bd \left(\frac{1}{\l_2} + \frac{Q_L\cdot Q_L}{|P_L\wedge Q_L|}\right) =0
.\!\!\!
\end{gather}

\subsection{Labelling the walls}\label{sec3.2}

The above formula (\ref{wallgamma}) for the walls of marginal stability might not look too complicated, but we should keep in mind that these are really inf\/initely many equations since $PGL(2,\Z)$ is not a f\/inite group.
Inspired by the fact that all the quantities involved in the above equation have simple transformation rules under the extended $S$-duality group, we would like to look for a~better way to organise the above equations of walls of marginal stability.

First note that the three-dimensional charge vector $\L_{P,Q} $ transforms as (\ref{action_PGL2Z})
under the $S$-duality transformation (\ref{extended_S-dual}).
Since the $S$-duality group leaves the Narain moduli $\m$ invariant, we conclude that the same transformation rule holds for the following vector
\[
\bem P_L \cdot P_L &  P_L\cdot Q_L \\  P_L\cdot Q_L &  Q_L\cdot Q_L \eem .
\]
Especially, the norm of the this vector which is proportional to $|P_L\wedge Q_L |$ is invariant under the transformation.

For the axion-dilaton, as familiar from the usual $S$-duality invariant formulation of the supergravity action, the
 following matrix also transforms under $PGL(2,\Z)$ in the same way as the charge vector
\[
\frac{1}{\l_2} \bem|\l|^2 & \l_1 \\ \l_1 & 1 \eem  .
\]

In particular, the following combination of these two vectors
\begin{gather*}
{\mathcal Z} = \frac{1}{\sqrt{|P_L \wedge Q_L|}} \bem P_L \cdot P_L &  P_L\cdot Q_L \\  P_L\cdot Q_L &  Q_L\cdot Q_L \eem + \frac{\sqrt{|P_L \wedge Q_L|}}{\l_2} \bem|\l|^2 & \l_1 \\ \l_1 & 1 \eem,
\end{gather*}
also transforms as
$
{\mathcal Z} \to \g({\mathcal Z} ) = \g {\mathcal Z} \g^T
$
under an extended $S$-duality transformation.

Using the vectors introduced above, the attractor equation can be written as
\[
\frac{1}{\l_2} \bem|\l|^2 & \l_1 \\ \l_1 & 1 \eem = \frac{1}{{|P_L \wedge Q_L|}} \bem P_L \cdot P_L &  P_L\cdot Q_L \\  P_L\cdot Q_L &  Q_L\cdot Q_L \eem
=  \frac{1}{{|P \wedge Q|}} \bem P \cdot P &  P\cdot Q \\  P\cdot Q &  Q\cdot Q \eem .
\]

As a side remark, let us note that the somewhat awkward-looking normalisation of the vector def\/ined above  gives the property that now the mass of a dyon, which involves $134$ moduli f\/ields,  is nothing but the norm of this single vector
\[
M_{P,Q}^2 =  |{\mathcal Z}|^2\;.
\]

As an attractor f\/low is a gradient f\/low of the BPS mass, we expect this vector to play a special role in the attractor f\/low.
But we are going to see in a moment that it is only the direction, not the length, of the vector ${\mathcal Z}$ which determines the existence or not of a certain two-centered solution. For convenience we will therefore def\/ine a ``unit vector'' $X$ by
\begin{gather}\label{def_X_vec}
X = \frac{{\mathcal Z} }{M_{P,Q}} =  \frac{{\mathcal Z} }{|{\mathcal Z}|}.
\end{gather}
Apparently this vector also transforms in the same way as ${\mathcal Z}$ under the $PGL(2,\Z)$ transformation.

With this notation, the walls of marginal stability for the $(P,0)$, $(0,Q)$ split of charges (\ref{simplewall}) lie on the co-dimensional one space characterised by the following equation{\samepage
\[
\big(X,\a_1 \big) = 0 ,
\]
and similarly for other splits (\ref{wallgamma}): $\big( X,\a \big) =0$ where $\a = \g^{-1}(\a_1)$.}

Now we have achieved the goal of organising the equations.
But to discuss in more details the relationship between the Weyl group of the Borcherds--Kac--Moody algebra and the physical walls of marginal stability, it is necessary to label these walls as well. As we mentioned before, the map (\ref{split_1}) between $PGL(2,\Z)$ elements and the split of charges into two $1/2$-BPS charges is not one-to-one. For example, the element which simply exchanges what we call the ``1st'' and the ``2nd'' decay products will not give a dif\/ferent split of charges. On the other hand, from the above expression for the walls of marginal stability, we see that the element $\g$ which gives $\g(\a_1)= \pm \a_1$ will not give a new physical wall. It can indeed be checked that these are also the elements which give the same split of charges when using the map (\ref{split_1}).
Using the fact that $PGL(2,\Z)$ is the group of symmetry for the root system of the Weyl group $W$ and by inspecting the action of the dihedral group (\ref{dihedral_action_1}), (\ref{dihedral_action_2}), it is not hard to convince oneself that the two-centered solutions with centers with $1/2$-BPS charges  discussed in the previous subsection are actually given by the positive real roots of the Borcherds--Kac--Moody algebra.

So we arrive at the conclusion that the relevant two-centered solutions are in one-to-one correspondence with the positive real roots of the Borcherds--Kac--Moody algebra, whose walls of marginal stability are given by
\begin{gather}\label{wall_2}
(X,\a) = 0  ,\qquad \a \in \D_+^{\rm re},
\end{gather}
where the set $\D_+^{\rm re}$ of positive real roots is def\/ined in (\ref{real_roots}).

\begin{figure}[t]
\centering
\includegraphics[width=6cm]{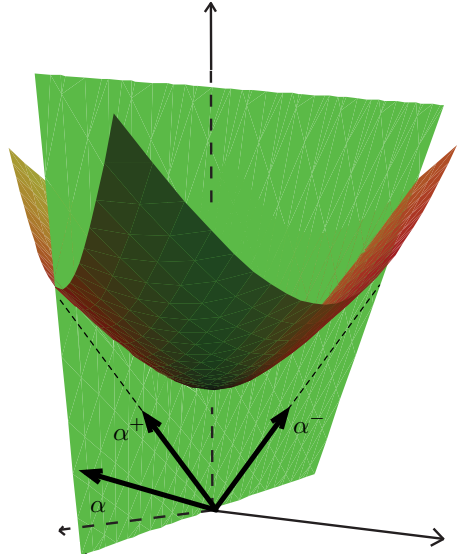} 
\caption{A plane $(X,\a)=0$ of orthogonality to a positive real root $\a$ always intersects the hyperboloid $|X|^2=1$, or equivalently the upper-half plane or the Poincar\'e disk. And the root $\a$ def\/ines two lightcone directions $\{\a^+$, $\a^-\}$ perpendicular to it, given by the intersection of the plane with the future light-cone.}\label{lightcone}
\end{figure}

The corresponding decay products, in terms of the positive real root $\a$, can be represented by the splitting of the vector $\L_{P,Q}$ of charge invariants as
\begin{gather}\nonumber
\L_{P_1,Q_1}  =  P_\a^2 \a^+  ,\qquad\L_{P_2,Q_2} =  Q_\a^2 \a^-,\\
\label{charge_split_new}
\L_{P,Q}  =  P_\a^2 \a^+ + Q_\a^2 \a^-  - |(P\cdot Q)_\a| \a ,
\end{gather}
where for a given $\a$, the set of the two vectors $\a^\pm$ is given by the requirement that (i) they are both lightlike and future-pointing and perpendicular to the root $\a$,
(ii) $\a^\pm/2$ lie in the weight lattice, namely that $\a^\pm$ as matrices have integral entries  (iii) they have inner product $(\a^+,\a^-)=-1 $. See Fig.~\ref{lightcone}.

These two-centered solutions have degeneracies
\begin{gather*}
 (-1)^{(P\cdot Q)_\a+1} |(P\cdot Q)_\a|d(P_\a)d(Q_\a),
\end{gather*}
which can again be shown by transforming the degeneracy (\ref{degeneracy_simple_split}) of the solution with charges $(P,0)$ and $(0,Q)$ by an $S$-duality transformation.

In particular, the quantities $P_\a^2$ and $Q_\a^2$, which are related to the ``oscillation level'' of the heterotic string and which determine the degeneracy of the $1/2$-BPS states, can be thought of as the ``af\/f\/ine length'' of the lightlike charge vector $\L_{P_1,Q_1}$ and $\L_{P_2,Q_2}$ of the decay products. Conversely, given the charges of the two centers, the walls of marginal stability are given by the requirement that the moduli vector $X$ is a linear combination of  the two charge vectors
$\L_{P_1,Q_1}$ and $\L_{P_2,Q_2}$.

More concretely, what the above formula (\ref{charge_split_new}) means is the following: given a spacelike vector, any future-pointing timelike vector can be split into a component parallel to it and a~future-pointing timelike vector perpendicular to it. And the latter can be further split into two future-pointing lightlike vectors lying on the plane perpendicular to the given spacelike vector. See Fig.~\ref{lightcone}. When the timelike vector is taken to be the original charge vector $\L_{P,Q}$ and the spacelike vector taken to be a positive real root, the two future-pointing lightlike vectors are then the charge vectors of the two decay products.

The meaning of the presence of a wall of marginal stability is that a BPS bound state of two particles
exists on one side of the wall and disappears when crossing into the other side. After deriving the location
of the walls for these bound states, we would like to know on which side these states are stable and on which side unstable. From the expectation that the attractor f\/low should resemble an RG-f\/low along which the number of degrees of freedom decreases, we expect the attractor moduli to always lie on the unstable side of the wall. This expectation is indeed correct. As mentioned earlier, the coordinate distance of the two centers is f\/ixed for given charges and moduli, a fact that can be understood as requiring the absence of closed time-like curve pathologies in the spacetime. More explicitly, it was analysed in \cite{Cheng:2007ch} that the coordinate distance between the two centers in the solution corresponding to the positive real root $\a$ (\ref{charge_split_new}) is given by
\begin{gather} \label{distance_dyon_center}
{| \vec{x}_{P_\a}-\vec{x}_{Q_\a}|} =  -\frac{1}{\sqrt{|P_L\wedge Q_L|}} \frac{(\L_{P,Q},\a)}{(X,\a)} ,
\end{gather}
with the normalization that the metric takes the standard form $ds^2 \to -dt^2 + d\vec{x}\,{}^2$ at spatial inf\/inity.

Since the distance between the two centers is always a positive number, one f\/inds that, in order for the bound state to exist,
the expression on the r.h.s.\ should better be positive as well. Hence the stability condition for the two-centered solution corresponding to $\a$ reads
\begin{gather}\label{stability_2}
(\L_{P,Q},\a)(X,\a) < 0,\qquad \a \in \D_+^{\rm re}.
\end{gather}

And indeed we see that the above stability can never be satisf\/ied at the attractor point of the moduli, where $X=\frac{\L_{P,Q}}{|\L_{P,Q}|}$. This is consistent with our intuition that no two-centered solutions with
a wall of marginal stability should exist at the attractor moduli.

\section[Weyl chambers and discrete attractor flow]{Weyl chambers and discrete attractor f\/low}\label{sec4}

Classically, a moduli space is a continuous space in which the vev's of the moduli f\/ields of the theory can take their values. A distinct path in this space  for a given superselection sector, namely given the total conserved charges, is the attractor f\/low of a single-centered black hole solution with a given starting point.

As we have established in last section, the moduli play a role in the spectrum of BPS states through the presence or absence of certain $1/4$-BPS bound states of two $1/2$-BPS objects. The walls of marginal stability for these bound states divide the moduli space into dif\/ferent regions in which the BPS spectrum is predicted to be constant by our supergravity analysis. This is because the spectrum only jumps when a wall of marginal stability is crossed, and for the purpose of studying the BPS spectrum of the theory we can identify the region bounded by a~set of walls to be a point.

From the above consideration, it is useful to consider a ``discrete attractor f\/low'', which brings one region in the moduli space to another. It is not dif\/f\/icult to see that such an operation forms a~group. Recall that in our theory,
the walls of marginal stability are in one-to-one correspondence with the positive roots of a Weyl group $W$, or the positive real roots of a~Borcherds--Kac--Moody algebra.  This implies that we should be able to identify the group of discrete attractor f\/low with the same Weyl group $W$. The aim of this section is to make this statement precise, and to address the implication of having such a group structure underlying the attractor f\/low.

\subsection{Weyl chamber and moduli space}\label{sec4.1}

To make the discussion more concrete let us visualise the situation. First recall that given a~point in the 134-dimensional moduli space, whether a given two-centered solution exists only depends on the combination of the moduli f\/ield encapsulated in the ``unit vector'' (\ref{def_X_vec}), which is future-pointing and which has norm $|X| =1 $.
First, the sheet of hyperbola of all future-pointing vector of f\/ixed norm $|X| =1 $ is equivalent to the upper-half (or Lobachevski) plane  by using the following map
\[
X= \frac{1}{\t_2} \bem  |\t|^2 & \t_1 \\ \t_1 & 1  \eem.
\]
One can therefore draw the walls $(X,\a)=0$ on the upper-half plane.
This representation of the walls on the upper-half  plane, however, does not really do justice to the symmetry of the problem. A nicer way to present the walls is to make a further map onto the Poincare disk by  the M\"obius transformation
\begin{gather*}
z=i\,\left(\frac{\t+e^{\frac{-\p i}{3}}}{\t+e^{\frac{\p i}{3}}}\right).
\end{gather*}

\begin{figure}[t]
\centering
\includegraphics[width=7.5cm]{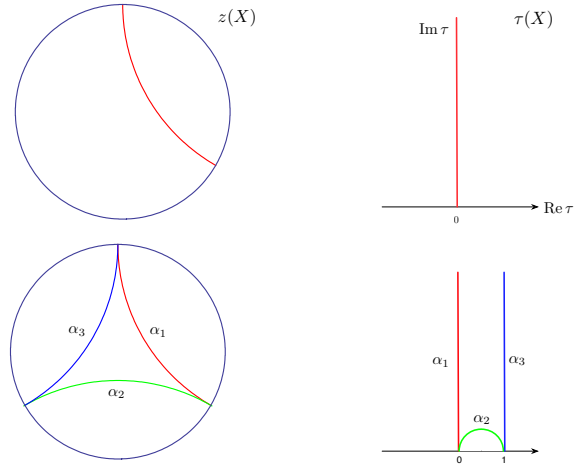} 
\caption{(i)
The wall of marginal stability for the two-centered solution with charges $(P,0)$ and $(0,Q)$,
projected onto the two-dimensional slice of moduli space equipped with a natural hyperbolic metric, and mapped to the Poincar\'e disk and the upper-half plane. (ii) The three simple walls $(X,\a_i)=0$.}\label{first_three_walls_graph}
\end{figure}

Now we are ready to draw the walls of marginal stability (\ref{wall_2}). First we will begin with $\a_1$, which corresponds to the two-centered solution with charges $(P,0)$, $(0,Q)$, for which the wall of marginal stability reads
$
(X,\a_1) = 0
$.
This gives an arc of a circle on the Poincar\'e disk, which is a geodesic with respect to the hyperbolic metric, and a straight line, or a degenerate circle, in the upper-half plane. See Fig.~\ref{first_three_walls_graph}.

As the next step, we will draw the walls given by other two roots $\a_2$ and $\a_3$ def\/ined in (\ref{3_simple_roots}). Notice that the three walls $(X,\a_i) = 0$, $i=1,2,3$ bound a triangle on the disk, and furthermore it is easy to show that the interior of the disk satisf\/ies $(X,\a_i)< 0 $. For example, the center of the triangle is given by the normalized version $\varrho/ |\varrho|$ of the Weyl vector
\begin{gather}
\label{weyl_vector_for_W}
\varrho =\frac{1}{2} (\a_1+\a_2+\a_3)=\bem1&1/2 \\ 1/2 &1\eem
\end{gather}
 satisfying $(\varrho,\a_i)=-1$.

As we will further justify in the next section, given a charge vector one should choose the simple roots such that $\L_{P,Q}$ lies inside the fundamental Weyl chamber. At this point this choice of simple roots can be seen as simply a convenient arbitrary choice.
For concreteness of the discussion and without loss of generality, we will now assume that such a choice is given by the three simple roots we used in section two, namely that the charge vector satisf\/ies
\[
 (\L_{P,Q},\a_i )  <   0 ,\qquad i =1,2,3,
\]
where $\a_i$ are the simple roots def\/ined in (\ref{3_simple_roots}).
In other words, from now on we will consider charges such that the vector $\L_{P,Q}$ lies  inside the large triangle in Fig.~\ref{first_three_walls_graph}. By the virtue of the relation between the Weyl group and the extended $S$-duality group $W \subset  PGL(2,\Z)$, we can always use the duality group to map a set of charges $(P,Q)$ to another set of charges for which the above is true.

\begin{figure}[t]
\centering
\includegraphics[width=10.5cm]{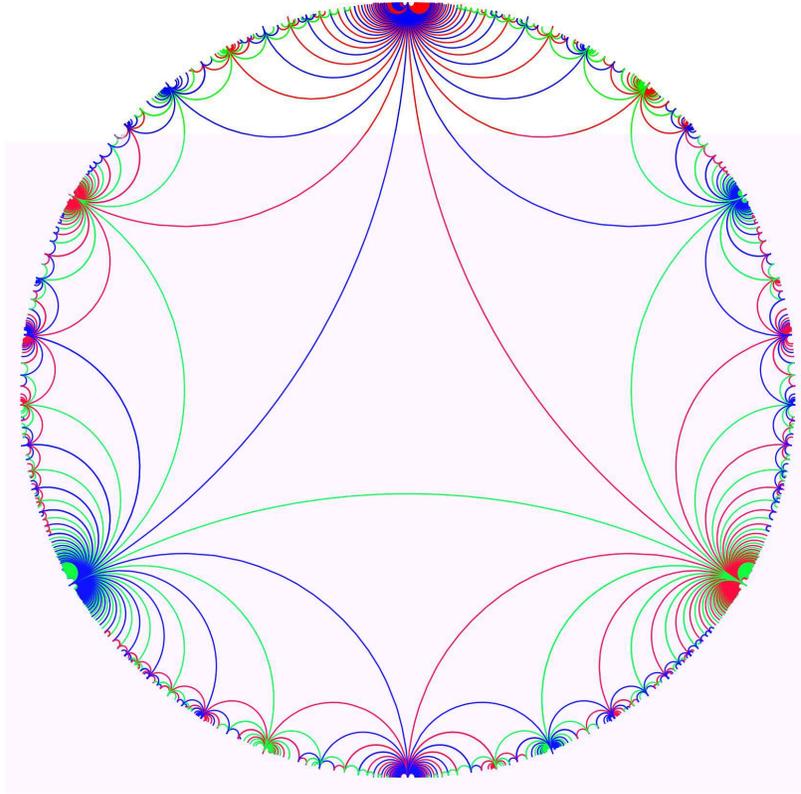} 
\caption{Tessellation of the Poincar\'e disk using the group $W$ generated by the ref\/lections with respect to the three ``mirrors'', namely the three sides of the regular triangle in the middle. Walls of the same color are mirror images of each other. Notice that each triangle has the same volume using the hyperbolic metric. The slight inhomogeneity of colours on the edge is an artifact of the computer algorithm we use.}\label{xxx}
\end{figure}

By def\/inition, all the other real roots, in particular all the other positive real roots, are related to these three simple roots by a Weyl ref\/lection $\a= w(\a_i)$. We can thus draw the rest of the walls of marginal stability
\[
(X,\a) = 0 ,\qquad \a \in \D_+^{\rm re}
\]
by ref\/lections of the triangle in Fig.~\ref{first_three_walls_graph} with respect to the three sides. This gives a tessellation of the Poincar\'e disk as shown in Fig.~\ref{xxx}. Notice that the f\/igure we draw can never be a truly faithful presentation of the real situation, since the group tessellates the disk with an inf\/inite number of triangles.

By def\/inition, the Weyl group divides the relevant part of the moduli space, namely the $X$-space, into dif\/ferent Weyl chambers bounded by the walls of orthogonalities with the positive roots. In other words, for any point in the moduli space, there exists a unique element of the Weyl group $w\in W$ such that the corresponding moduli vector $X$ lies in the Weyl chamber
\[
X \in w({\cal W}) \Leftrightarrow (X,w(\a_i)) < 0 .
\]

Because these walls, or mirrors, of the Weyl chambers are exactly the physical walls of marginal stability  corresponding to the split into two centers (\ref{charge_split_new}),
we conclude that the BPS spectrum does not jump when the moduli change within a given triangle. In other words,
the Weyl chambers are precisely the regions in the moduli space in which the BPS spectrum stays constant, and there is a dif\/ferent dyon degeneracy associated to each Weyl chamber $w({\cal W})$.

\subsection{A hierarchy of decay}\label{sec4.2}

Now we would like to discuss what the group structure of the moduli space described above implies for the dyon spectrum at a given moduli. In principle, given a point in the moduli space we know exactly which two-centered solutions exist by using the stability condition (\ref{stability_2}) we worked out in the last section. But in fact we know less than it might seem. This is because there are inf\/initely many decay channels, or to say that there are inf\/initely many real roots in the Borcherds--Kac--Moody algebra, so given a moduli vector $X$, even equipped with the stability condition one is never able to give a list of two-centered solutions within f\/inite computation time. In this subsection we will see how the group structure changes this grim outlook.

First note that, in the central triangle, namely the fundamental Weyl chamber, there is no two-centered bound states. This can be seen as follows. Recall that we have chosen the simple roots for  given charges such that
\[
(\L_{P,Q},\a_i) < 0 ,\qquad i=1,2,3,
\]
which implies
\[
(\L_{P,Q},\a) < 0 ,\qquad \text{for all}\  \ \a \in \D_+^{\rm re}
\]
simply from the def\/inition of the positive real roots. Now the same thing holds for a point $X\in {\cal W}$ inside the fundamental Weyl chamber, namely
$
(X,\a) < 0 $  for all $\a \in \D_+^{\rm re}$,
hence we see that the stability condition (\ref{stability_2}) will never be satisf\/ied for any split of charges.

We therefore conclude that the fundamental Weyl chamber represents the ``attractor region'' in the moduli space, namely the same region (chamber) as where the attractor point lies and in which none of the relevant two-centered solutions exists.

The absence of two-centered solutions is of course not true anymore once we move out of the fundamental Weyl chamber. Consider for example the neighbouring Weyl chamber $s_1({\cal W})$, obtained by ref\/lecting the fundamental chamber with respect to one of the simple roots $\a_1$. Because this ref\/lection takes $\a_1 \to -\a_1$ and permutes the rest of the positive real roots (see Theorem~\ref{length_posi_negi}), we conclude that
\[
 (X,\a_1) >0
 \] and
\[
(X,\a)< 0 \qquad\text{for all}\  \ X\in s_1({\cal W}), \ \ \a\neq \a_1, \ \ \a \in \D_+^{\rm re},
\]
which is also obvious from the picture.
This means that there is now one two-centered solution corresponding to the split into charges $(P,0)$, $(0,Q)$ and no other ones.

We can now go on with this process to every Weyl chamber in the  moduli space: go to the next-neighbouring chamber, and the next-next-neighbouring, and so on, with the condition that the path doesn't walk ``backwards'', or more precisely that the length function (see def\/inition~(\ref{length_function})) of the corresponding group element always increases.

In general, considering an arbitrary point on the disk
\[
X \in w({\cal W})
\]
for some $w\in W$. For any group element, we can always decompose it as the following ``shortest word'' in terms of the three group generators
\begin{gather}\nonumber
w = s_{i_1}s_{i_{2}}\cdots s_{i_n},\\
\label{decomposition_1}
 i_m \in \{1,2,3\},\quad  i_m \neq i_{m\pm 1} \quad \text{for all} \ \ m = 2,\dots,n-1.
\end{gather}
From the fact that a ref\/lection with respect to the root $w(\a_i)$ is given by the group ele\-ment~$w s_i w^{-1}$, we can describe the Weyl chamber $w({\cal W})$ as given by the following successive ref\/lections of the fundamental Weyl chamber:
\begin{gather}\label{bruhat_decay}
w_0 = {\boldsymbol 1}  \xrightarrow[\a_{i_1} ]{}  \,w_{1} \,\xrightarrow[w_{1}(\a_{i_{2}}) ]{}\, w_{2} \;\cdots \; \xrightarrow[w_{n-1} (\a_{i_n})]{}  \quad w_n=w,
\end{gather}
where `` $\xrightarrow[\;\;\a\;\; ]{}$'' means ``ref\/lecting with respect to the root $\a$''
and the intermediate group ele\-ments~$w_m$ are given by
\begin{gather*}
w_m = s_{i_1}\cdots s_{i_m} ,\qquad m\leq n .
\end{gather*}

In other words, for a given point in the moduli space and its Weyl chamber
$
X\in w({\cal W}) ,
$
by following the above path from the attractor region ${\cal W}$, or the fundamental Weyl chamber, to~$w({\cal W})$, we can read out the two-centered solutions which exist at the point $X \in {\cal W}$. These are given by the charge splitting (\ref{charge_split_new}) with now
\begin{gather}\label{death_row}
\a \in \bigl\{ w_{m-1}(\a_{i_m}),\  m=1,\dots,n\bigr\} \subset \D_{+}^{\rm re}.
\end{gather}

 In other words, when we follow the journey from the attractor region ${\cal W}$ to the Weyl cham\-ber~$w({\cal W})$ where the moduli is, namely when we follow the inverse attractor f\/low to the point in the moduli space under consideration, we will successively cross the walls of marginal stability corresponding to the roots $\a_{i_{1}}$, $w_{1}(\a_{i_{2}})$, \dots, and f\/inally $w_{n-1}(\a_{i_{n}})$.
 This gives a simple inventory to read out the list of two-centered solutions for any given point in the moduli space, provided that we know the shortest decomposition of the group element $w$ in terms of a string of generators (``letters'').

To complete the construction of the inventory, we also give a very simple algorithm to determine such a decomposition given an arbitrary lightlike, future-pointing vector $X$. Given a~point $X$, we would like to determine the shortest-length string  (\ref{decomposition_1}),
such that its correspon\-ding chain of ref\/lections (\ref{bruhat_decay}) induces the following successive mappings of the vector $X$ into the fundamental Weyl chamber
\begin{gather}\label{bruhat_decay_2}
 X_0 \in {\cal W}  \xrightarrow[\a_{i_1} ]{}  \,X_1\,
 \xrightarrow[w_{1}(\a_{i_{2}}) ]{} \, \cdots\, \xrightarrow[w_{n-1} (\a_{i_n})]{}  X_n=X \in w({\cal W}) .
\end{gather}
One can show that the string is determined as follows: suppose
\begin{gather*}
X_m = s_{i_1}\cdots  s_{i_m} X_0 ,\qquad m\leq n ,
\end{gather*}
then  $i_m\in\{1,2,3\}$ is given by the condition
\begin{gather*}
(\a_{i_m},X_m) > 0 ,
\end{gather*}
which has at most one solution for $i_m$. We can go on with this process for $X_{m-1}$ until the above equation has no solution anymore, corresponding to when the three expansion coef\/f\/icients $X_0^{(i)}$ of $X_{0} = \sum_{i=1}^3 \a_i X_0^{(i)} $ satisfy the triangle inequality. This is when the decay ends and when the moduli have f\/lown to the attractor region given by the fundamental Weyl chamber.

Notice that there is a hierarchy of decay in this process. Namely, considering another point in the moduli space which is in the Weyl chamber $w_m({\cal W})$ with $m< n$, applying the above argument  shows that the two-centered solutions existing in that Weyl chamber are given by the f\/irst $m$ positive roots in the above list. Specif\/ically, this argument shows that there is nowhere in the moduli space where the bound states given by the root $w_{n-1}(\a_{i_{n}})$ exists without all the other $n-1$ bound states given by $w_{m-1}(\a_{i_{m}})$, $m<n $ in front of it in the row.

 This hierarchy among two-centered bound states clearly stems from the hierarchy among elements of the group $W$. Indeed, the ordering in (\ref{bruhat_decay}) is an example of what is called the ``weak Bruhat order'' $w_m < w_{m+1}$ among elements of a Coxeter group. See (\ref{weak_order}) for the def\/inition. This ordering has made an earlier appearance in theoretical physics as giving the proposed arrow of time in a dif\/ferent context in \cite{Fre:2007hd}.

\begin{figure}[t]
\centering
\includegraphics[width=10.0cm]{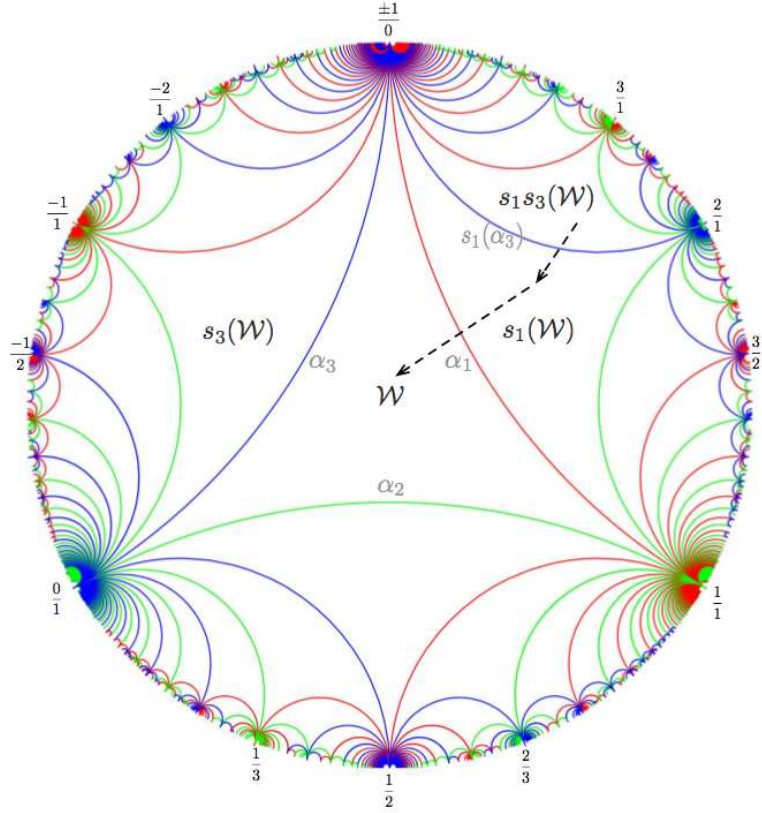} 
\caption{(i) An example of a discrete attractor f\/low from $X \in s_1s_3({\cal W}) $ to the attractor region
${\cal W}$, passing through two walls of marginal stability $(s_1(\a_3),X)=0$ and $(\a_1,X)=0$.  (ii) The boundary of the disk can be identif\/ied with the boundary of the light-cone, and in turn be identif\/ied with a compactif\/ied real line using the map (\ref{map_lightcone_number}). In this way there is a pair of rational numbers associated to each positive real roots, and a Weyl chamber is given by such a pair with its mediant. Furthermore, the discrete attractor f\/low can be thought of as a process of ``coarse-graining'' the rational numbers.}\label{disk2}
\end{figure}

In our case this weak Bruhat ordering has an interpretation reminiscent of the RG-f\/low of the system. From the integrability condition (\ref{distance_dyon_center}),
now with the two-centered solutions given by (\ref{death_row}),
we see that the ordering of the decay is exactly the ordering of the coordinate size of the bound state. In other words, roughly speaking, the ordering we discussed above can be summarised as the principle that the bound state which is the bigger in size are more prone to decay than the smaller ones, which is a fact reminiscent of the usual RG-f\/low phenomenon. See Fig.~\ref{disk2} for a simple example of the f\/low beginning from a point in $s_1s_3({\cal W})$.

As was alluded to earlier, we can identify the path taken by such a string of ref\/lections as the path taken by a discrete version of attractor f\/lows. In other words, if we identify all points in a given Weyl chamber, justif\/ied by the fact that all points there have the same BPS spectrum, the usual continuous attractor f\/low reduces to the successive ref\/lections discussed above. To argue this, note that the (single-centered) attractor f\/low also gives such a structure of hierarchy among multi-centered solutions, because an attractor f\/low can only cross a given wall at most once. Using the ${\cal N}=2$ language, this is because for a single-centered solution, the quantity $\im(\bar{Z} Z')/|Z|$ is linear in ${1}/{r}$, the inverse of the coordinate distance from the black hole, where $Z'$ is the central charge of another arbitrary charge at the moduli where the f\/low is, and can therefore only pass zero at most once \cite{Denef:2001xn}.  On the other hand, such a hierarchy among solutions can only be a property of the structure of the walls in moduli space, since the stability condition is a local condition on the moduli and is in particular path-independent. Hence we have to conclude that the group and the f\/low must cross the walls in the same order. This justif\/ies our claim that the group is simply the discretised group of attraction.

\subsection[Arithmetic attractor flows]{Arithmetic attractor f\/lows}\label{sec4.3}

In this section we will discuss the arithmetic aspects of our newly def\/ined discrete attractor f\/low  group. We will see how from this point of view, the attractor f\/low can be thought of a process of coarse-graining the rational numbers.

First let us again review the equation (\ref{charge_split_new}) for the splitting of charges into two $1/2$-BPS centers, which states that for any positive real root $\a$, the charge vector $\L_{P_1,Q_1}$, $\L_{P_2,Q_2}$ of the associated two-centered solution are given by the component of the total charge vector $\L_{P,Q}$ along the two lightcone directions  perpendicular to $\a$.

From the def\/initions of the light-like vectors $\a^\pm$, it is not dif\/f\/icult to see that the two-centered solutions can equivalently be given by a set of a pair of rational numbers $\{b/a,d/c\}$ satisfying $ad-bc=1$,
such that the lightlike vectors are given by
\begin{gather}\label{map_lightcone_number}
\{\a^+,\a^-\} = \left\{ \bem b^2 & ab \\ ab & a^2\eem, \bem d^2 & cd \\ cd & c^2\eem \right\}.
\end{gather}
Notice that exchanging the two rational numbers amounts to exchanging $(P_1,Q_1)$ and $(P_2,Q_2)$, which obviously does not give a new solution.
Without loss of generality, we now impose that $a\geq 0$, $c\geq 0$,
while $b$, $d$ can take any sign.

To be more precise, given such a pair of rational numbers, the corresponding positive roots~is
\begin{gather*}
\a = \bem 2bd & ad+bc \\ ad+bc & 2ac \eem ,
\end{gather*}
and the corresponding charge splitting is (\ref{split_1})
\begin{gather}\label{charge_split_rational_number}
\bem P_1 \\ Q_1 \eem = (-c P + dQ) \bem b \\ a \eem  ,\qquad
\bem P_2 \\ Q_2 \eem = (a P - bQ) \bem d \\ c \eem .
\end{gather}

In other words, the above formula gives an alternative labelling of the two-centered solutions of the theory by a pair of rational numbers
\begin{gather}\label{pair_rational_number}
\left\{\frac{b}{a},\frac{d}{c}\right\} ,\qquad ad-bc=1 ,\qquad a,c \geq 0 .
\end{gather}

In particular, the three simple roots (\ref{3_simple_roots}) correspond to the three sets of rational numbers $\{\frac{-1}{0},\frac{0}{1}\}$,  $\{\frac{0}{1},\frac{1}{1}\}$,  $\{\frac{1}{1},\frac{1}{0}\}$ respectively.

Now we would like to know what the discrete attractor f\/low, def\/ined in the last subsection, looks like in terms of the presentation in terms of rational numbers. From Fig.~\ref{disk2} it is obvious that, for a positive root bounding the Weyl chamber $w({\cal W})$, one of the following two roots $wOw^{-1}(\a)$ and $wO^2w^{-1}(\a)$ must be negative, where $O$ is the order three generators of the dihedral group corresponding to a rotation of $120^\circ$ (\ref{dihedral_action_2}), and this must be the root with respect to which the last ref\/lection in the sequence  (\ref{bruhat_decay}) is. In equations and in the notation of last subsection, this root of last ref\/lection is given by $w_{n-1}(\a_{i_n}) = -w(\a_{i_n})$ where the positive root~$\a$ is given by $\a=w(\a_i)$ with some $i\neq i_n$.

Therefore, from the computation which gives the following expression for $-wOw^{-1}(\a)$ and $-wO^2w^{-1}(\a)$
\[
\bem 2d(b-d) & bc+ad-2cd \\ bc+ad-2cd & 2a(a-c) \eem  ,\qquad \bem 2b(d-b) & bc+ad-2ab \\ bc+ad-2ab & 2c(c-a) \eem ,
\]
we conclude that given a two-centered solution corresponding  to the pair rational numbers $\{b/a$, $d/c\}$ with $a,c \geq 0$ and $ad-bc=1$, we can read out the next two-centered solution on the list of decadence as another pair of rational numbers
\[
\left\{\frac{b}{a},\frac{d-b}{c-a}\right\}\quad \text{if}\ \ c\geq a, \ bd \geq b^2,\qquad
\left\{\frac{d}{c},\frac{b-d}{a-c}\right\}\quad \text{if}\ \ a \geq c, \ bd \geq d^2.
\]

This rule is actually much simpler than it might seem. Consider the boundary of the Poincar\'e disk as a compactif\/ied real line, namely with $\frac{\pm1}{0}$ identif\/ied, then the map (\ref{charge_split_rational_number}), (\ref{pair_rational_number}) associates with each wall of a Weyl chamber, or equivalently each wall of marginal stability, a pair of rational numbers satisfying the relevant conditions. See Fig.~\ref{disk2}. Now what we saw above simply means: f\/irstly, any triangle is bounded by a set of three rational numbers of the form
\begin{gather*}
\left\{\frac{b}{a},\frac{b+d}{a+c},\frac{d}{c}\right\},\qquad ad-bc=1,\qquad a,c \geq 0.
\end{gather*}
The middle element $(b+d)/(a+c)$ is called the ``mediant'' of the other two, and can be easily shown to always lie between the two
\[
\frac{b}{a}<\frac{b+d}{a+c}<\frac{d}{c}.
\]
Secondly, the direction of the attractor f\/low is that it always f\/lows to the numbers with smaller $|a|+|b|$ and $|c|+|d|$, and can therefore be seen as a f\/low of coarse-graining the rational numbers.

\begin{figure}[t]
\centering
\includegraphics[width=12cm]{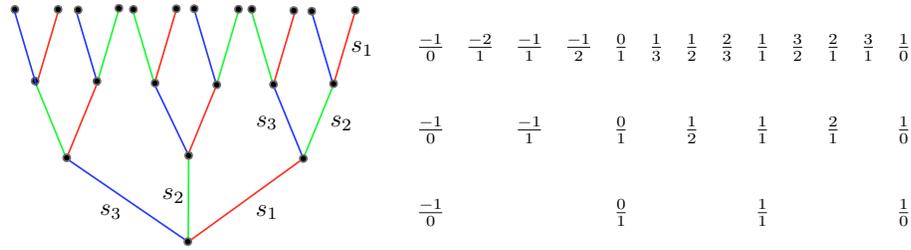} 
\caption{(l) The (f\/irst three levels of the) weak Bruhat ordering of the group $W$, which corresponds to the hierarchy of the wall-crossing of the theory. (r) The corresponding coarse-graining of the rational numbers, which is a part of the Stern--Brocot tree generalised to the whole real line.}\label{stern_brocot_bruhat}
\end{figure}

As an illustration of the above statement, in Fig.~\ref{stern_brocot_bruhat} we show the rational numbers correspon\-ding to the f\/irst three levels of inverse attractor ref\/lections from the attractor region, corresponding to the f\/irst three levels of the weak Bruhat tree.

Note that this is simply the generalisation of the so-called Stern--Brocot tree to the negative part of the real line, and notice that {\it all} the rationals are contained in this tree. In particular, the Farey series is contained in the middle part of the tree.

\section{The algebra and the BPS states}\label{sec5}

So far we have seen how a rank three hyperbolic Coxeter (or Weyl) group underlies the structure of wall-crossing of the present theory. In particular, we have seen that the vectors in its root lattice can be identif\/ied as the vectors of $T$-duality invariants of the dyons. Furthermore, this group has the physical interpretation as the group of crossing the walls of marginal stability, or equivalently the group of discretized attractor f\/lows.

As we have mentioned in the introduction, the dyon-counting generating function coincides with the square of the denominator of a certain Borcherds--Kac--Moody algebra, and the group we have been discussing so far coincides with the Weyl group of this  Borcherds--Kac--Moody algebra. In particular, the set of roots of the Weyl group coincides with the set of {\it real} roots of this algebra.

But apart from this observation, so far we have not really used the structure of this Borcherds--Kac--Moody algebra. Indeed what we have not discussed is how the group structure of wall-crossing implies for the dyon degeneracies in relation to the BPS algebra. Notice that the same Coxeter group can  as well be the Weyl group of other generalized Kac--Moody algebras, for example an Lorentzian Kac--Moody algebra without imaginary simple roots similar to the ones described in \cite{Feingold-Frenekel,Feingold-Nicolai}. To understand the physical role of the algebra, it is therefore not suf\/f\/icient to understand only the role played of its Weyl group.

In this section we will therefore go beyond the Weyl group and discuss how the dyon-counting formula can be interpreted as relating the degeneracy of dyons to the dimensionality of some weight space. In particular, given a super-selection sector of f\/ixed total charges, we will see how the relevant representation of the algebra changes when the background moduli change.

\subsection[Borcherds-Kac-Moody algebra and the denominator formula]{Borcherds--Kac--Moody algebra and the denominator formula}\label{sec5.1}

First we will begin with a short review of the Borcherds--Kac--Moody superalgebras and its (Weyl--Kac--Borcherds) denominator formula, and how the dyon-counting generating func\-tion $\F(\O)$ at the r.h.s.\ of (\ref{DVV0}) can be expressed in terms of the data of the algebra.

A Borcherds--Kac--Moody superalgebra is a generalisation of the usual Lie algebra by the following facts: (i) the Cartan matrix is no longer positive-def\/inite (``Kac--Moody''), (ii) there are also the so-called ``imaginary'' simple roots with lightlike or timelike length (``generalised'' or ``Borcherds''), (iii) it is $\Z_2$-graded into the ``bosonic'' and the ``fermionic'' part (``super''). We will summarise the important properties of these algebras that we will use later. See~\cite{Ray} or the appendix of \cite{GN1} for a more systematic treatment of the subject.

Consider a set of simple roots $_s\D$ and a subset of fermionic (bosonic) simple roots $_s\D_{\bar{1}}$ ($_s\D_{\bar{0}}$). A  generalised Cartan matrix is a real matrix of inner product $(\a,\a')$ with $\a,\a'\in  {_s}\D$ that satisf\/ies the property
\[
 2\frac{(\a,\a')}{(\a,\a)} \in \Z\quad   \text{if}\ \ (\a,\a)>0.
 \]
Furthermore we will restrict our attention to the special case of Borcherds--Kac--Moody algebra without odd real simple roots, which means
\[
 (\a,\a)\leq 0 \quad\text{if}\ \ \a\in {_s}\D_{\bar{1}}.
 \]
We will say that $\a \in {_s\D}$ is a real simple root if $(\a,\a)>0$ and an imaginary simple root otherwise.

Then the Borcherds--Kac--Moody superalgebra ${\mathfrak{g}}(A,S)$ is the Lie superalgebra with even generators $h_\a$ of the Cartan subalgebra, and other even generators $e_\a$, $f_\a$ with $\a \in$$_s\D_{\bar{0}}$ and odd generators  $e_a$, $f_a$ with $i \in$$_s\D_{\bar{1}}$, satisfying the following def\/ining relations
\begin{gather*}
[e_\a,f_\a'] =  h_\a \quad \text{if} \ \a=\a', \qquad \text{and zero otherwise},\\
[h_\a,h_\a']=0,\\
[h_\a,e_{\a'}] = (\a,\a') e_{\a'},\qquad [h_\a,f_{\a'}] =- (\a,\a') f_{\a'},\\
(\text{ad}\, e_\a)^{1-2\frac{(\a,\a')}{(\a,\a)}}e_{\a'} =(\text{ad}\, f_\a)^{1-2\frac{(\a,\a')}{(\a,\a)}}f_{\a'}= 0 \quad\text{if} \ \a\neq \a', \qquad (\a,\a)>0,\\
 [e_\a,e_{\a'}] =   [f_\a,f_{\a'}] = 0\quad \text{if} \ (\a,\a')=0.
\end{gather*}

Another important concept we need is that of the root space. For later use we have to introduce more terminologies. The {\it root lattice} $\G$ is the lattice (the free Abelian group) generated by $\a$, $\a \in$$_s\D$, with a real bilinear form $(\a,\a')$. The Lie superalgebra is graded by $\G$ by letting $h_\a$, $e_\a$, $f_\a$ have degree $0$, $\a$ and $-\a$ respectively. Then a vector $\a\in \G$ is called a {\it root} if there exists an element of ${\mathfrak{g}}$ with degree $\a$. A root $\a$ is called {\it simple} if $\a \in$$_s\D$, {\it real} if it's spacelike $(\a,\a) > 0$ and {\it imaginary} otherwise. It is called  {\it positive (negative)} if it is a positive- (negative-) semi-def\/inite linear combinations of the simple roots. It can be shown that a root is either positive or negative. Furthermore, the {\it Weyl group} $W$ of  ${\mathfrak{g}}$ is the group generated by the ref\/lections in $\G\otimes \R$ with respect to all real simple roots. A Weyl vector $\varrho$ is the vector with the property
\[
(\varrho,\a) = -\frac{1}{2} (\a,\a)\quad\text{for all simple real roots}\ \a.
\]
It is easy to check that, for the algebra at hand with the choice of real simple roots as discussed in Section~\ref{sec2}, the vector (\ref{weyl_vector_for_W}) is indeed the Weyl vector satisfying the above condition.

Just as for ordinary Kac--Moody algebras, there is the following so-called denominator formula for the Borcherds--Kac--Moody algebras
\begin{gather}\label{denominator_general}
e(-\varrho) \prod_{\a \in \D_+} \big( 1- e(-\a) \big)^{\text{mult}\, \a} = \sum_{w\in W}\,\e(w) w(e(-\varrho) \,S),
\end{gather}
where $\D_+$ is the set of all positive roots, $\e(w) = (-1)^{\ell(w)}$ where $\ell(w)$ is the length of the word~$w$ in terms of the number of generators, def\/ined in (\ref{length_function})\footnote{Warning: the conventions of the signs of the above formula, in particular the signs of the Weyl vector, do vary in the existing literature.}. The $e(\m)$'s are formal exponentials satisfying the multiplication rule $e(\m)e(\m')=e(\m+\m')$. There are dif\/ferences between this product formula and the usual Weyl denominator formula due to, f\/irst of all, the fact that it's ``super''. Concretely, we have used the following def\/inition for the ``multiplicity'' of roots ${\rm mult}\,\a$:
\[
{\text{mult}}\, \a = \dim {\mathfrak{g}}_{\a,\bar{0}}  -\dim {\mathfrak{g}}_{\a,\bar{1}}.
\]
Furthermore, there is a correction term $S$ on the right-hand side of the formula due to the presence of the imaginary roots. The exact expression for $S$ is rather complicated for generic Borcherds--Kac--Moody superalgebras and can be found in~\cite{Ray,GN1}.

Now we will concentrate on the algebra under consideration.
The dyon-counting generating function $\F(\O)$ in (\ref{DVV0})  proposed in \cite{Dijkgraaf:1996it} can be written in the following form
\begin{gather}\label{DVVPF_2}
\F(\O)^{1/2} = e^{-\p i(\varrho,\O)}\,\prod_{\a\in  \D_+}\,\big(1-e^{-\p i (\a,\O)}\big)^{\frac{1}{2}c(|\a|^2)},
\end{gather}
where the product is taken over the integral vector
\begin{gather*}
\a \in \bigl\{ \Z_+\a_1+\Z_+\a_2+\Z_+\a_3\bigr\}
\end{gather*}
and $\varrho$ is the Weyl vector.

This is precisely in the form of the l.h.s.\ of the denominator formula
(\ref{denominator_general}), with the identif\/ication
\begin{gather}\label{multiplicity}
\text{mult}\, \a= \frac{1}{2} c (  |\a|^2),
\end{gather}
where $c(k)$ is the $k$-th Fourier coef\/f\/icient of the $K3$ elliptic genus \cite{Eguchi:1988vra}, namely
\begin{gather*}
\chi(\s,\n) = \sum_{n\geq 0, \ell \in \Z}
c(4n-\ell^2) q^n y^\ell,\qquad q=e^{2\p i \s}, \qquad y=e^{-2\p i \n},
\end{gather*}
with $c(-1)=2$, $c(0)=20$.

 As $c(k)$ grows with $k$, to ensure the convergence of the above function $\O$ should be restricted to lie in the Siegel upper-half plane, obtained by complexifying the vector space $\R^{2,1}$ introduced before and taking only the future light-cone for the imaginary part
\[
\O \in \R^{2,1}+ iV^+.
\]

From the above-mentioned relation between the $K3$ elliptic genus and the function $\F(\O)$, one can show that $\F(\O)$ is an automorphic form of weight ten of the group $Sp(2,\Z)$ \cite{Borcherds1,GN1,Kawai:1995hy,Harvey:1995fq}. In particular, it is invariant under the transformation
$
\O \to \g(\O)
$
for all elements $\g$ of the extended $S$-duality group $PGL(2,\Z)$.

From the transformation property of the above automorphic form under the Weyl group $W$ introduced in Section~\ref{sec2}, we can also rewrite it in a form as the right-hand side of (\ref{denominator_general}). From this equivalence one can therefore read out the set of even and odd imaginary simple roots and therefore construct a ``automorphic-form corrected'' Borcherds--Kac--Moody superalgebra, whose denominator is the Siegel modular form $
\F(\O)^{1/2}$ of weight f\/ive
and has the (graded) multiplicities of the roots given by the Fourier coef\/f\/icient of the $K3$ elliptic genus as (\ref{multiplicity})  \cite{GN1}. In particular, the set $\D_+$ can now be interpreted as the set of all positive roots. Notice that the property that $c(k)=0$ for $k<-1$ is consistent with the claim that all the real roots, namely the space-like $\a$ with $c(-(\a,\a)/2)$ non-zero are the images of the three vector $\a_{1,2,3}$ introduced in Section~\ref{sec2} under the Weyl group. Indeed it is easy to check that the set of real roots $\D^{\rm re}$ of the Borcherds--Kac--Moody algebra def\/ined above coincides with the set of all roots of the Weyl group $W$.

For later use we will also review some properties of a class of  poles of the above partition function. This property of the automorphic form $\F(\O)$ can be best understood as a special property of the Fourier coef\/f\/icients of the $K3$ elliptic genus. Recall that when we put the parameter $\n=0$, the elliptic genus reduces to the Witten index of the conformal f\/ield theory, which is given by the Euler number of the target manifold and equals $24$ in the present case of~$K3$ manifold. In other words
\begin{gather*}\nonumber
\chi(\s,\n=0) = \left( 20 + 2 (y+y^{-1}) + \sum_{n>0, \ell \in \Z}
c(4n-\ell^2) q^n y^\ell \right) \Big\lvert_{y=1} =24\\
\phantom{\chi(\s,\n=0)}{} \Rightarrow   \sum_{ \ell \in \Z}c(4n-\ell^2)  = 0 \qquad\text{for all}\quad
n>0.
\end{gather*}

One can check that this property of the integers $c(4n-\ell^2)$ translates into the following expression for the pole of the automorphic form $\F(\O)$
\[
\frac{\F(\O)}{(y^{1/2}-y^{-1/2})^2}\Big\lvert_{y\to1} = \eta^{24}(\r)   \eta^{24}(\s) .
\]

Employing the invariance of $\F(\O)$ under the symmetry group $PGL(2,\Z)$ of the root system, this can be generalized into
\begin{gather}\label{pole_look_like}
\frac{\F(\O)}{(e^{-\p i(\O,\a)/2} - e^{\p i(\O,\a)/2})^2}\Big\lvert_{(\O,\a)\to 0} =
\eta^{24}(-2\p i (\a^+,\O)) \eta^{24}(-2\p i(\a^-,\O))
\end{gather}
which holds for arbitrary positive real root $\a$ and the associated $\a^\pm$ are def\/ined as in (\ref{charge_split_new})\footnote{This equation might be thought of as a ``automorphized'' version of the G\"ottsche's generating function of the Euler characteristic of the Hilbert scheme of $m$ points on the manifold $M$ \cite{Gottsche,Dijkgraaf:1996xw}
\[
\sum_{m=0}^\inf\, p^m \chi(\text{Sym}^mM) = \prod_{k=1}^\inf \left(\frac{1}{1-p^k}\right)^{\chi(M)} ,
\]
where the manifold $M$ is the $K3$ manifold in this case.}.

As was studied in \cite{Cheng:2007ch}, this property makes possible
that the wall-crossing phenomenon can be incorporated in the dyon-counting integral by a moduli-dependent contour prescription.
In Section~\ref{sec5.3} we show how it is possible to derive the wall-crossing formula microscopically using the above property of the denominator formula and the dif\/ferent representations of the Borcherds--Kac--Moody algebra at dif\/ferent point in the moduli space proposed in Section~\ref{sec5.2}.

\subsection{Wall-crossing and the representation of the algebra}\label{sec5.2}

The fact that the dyons are counted with a generating function which is simply the square of the denominator formula of a generalized Kac--Moody algebra strongly suggests a physical relevance of this superalgebra in the BPS sector of the theory. In the previous sections we have explained the physical relevance of the real roots, namely that the walls of marginal stability in supergravity are labelled by positive real roots. In this subsection we will see how the imaginary roots enter the counting of dyonic states. Nevertheless we will not see their appearance as directly in the low-energy supergravity theory as the real roots.

To extract the actual dyon degeneracies from the generating function (\ref{DVVPF_2}), notice that in the integral (\ref{DVV0}) we have not specif\/ied along which contour the integration  should be performed.
This would not have been a problem if the answer of the integral does not depend on the choice of a contour of integration.
But in \cite{Sen:2007vb,Dabholkar:2007vk} it was pointed out that it is not the case due to the intricate structure of the poles of the integrand. Also the l.h.s.\ of (\ref{DVV0}) is ambiguous: from the discussion in the previous sections we have seen that the BPS degeneracies depend on the background moduli due to the existence of walls of marginal stability in the moduli space.

Nevertheless, this seemingly embarrassing situation is actually a bless in disguise. It was shown that the moduli dependence of the spectrum can be incorporated in the above dyon-counting formula through a moduli-dependent choice of contour \cite{Cheng:2007ch}. In its full glory the dyon-counting formula now reads
\begin{gather}\label{DVV_integral_2}
(-1)^{P\cdot Q+1} D(P,Q)\lvert_{\l,\m} = \oint_{{\cal C}(P,Q)\lvert_{\l,\m}}   d\O  \left( \frac{e^{i\frac{\p}{2}  (\L_{P,Q},\O) }}{e^{-\p i(\varrho,\O)}\prod\limits_{\a\in \D^+}\big( 1-e^{-\p i(\a,\O)}\big)^{\text{mult}\,\a}} \right)^2 ,
\end{gather}
 where $D(P,Q)\lvert_{\l,\m}$ is the degeneracy of BPS states with total charge $(P,Q)$ and evaluated at background moduli $\l$ (axion-dilaton) and $\m$ (Narain moduli). And the contour of integration is given by
 \[
{{\cal C}(P,Q)\lvert_{\l,\m}}  = \big\{\re\, \O \in T^3 , \im\,\O =  \varepsilon^{-1} X\big\} ,
 \]
 where $\epsilon\ll 1$ is any small positive real number which plays the role of a regulator.

Now we would like to pause and ask ourselves the following question: what does it mean that a dif\/ferent choice of contour gives a dif\/ferent answer for the BPS degeneracy? After all, the dyon-counting partition function (\ref{DVVPF_2}), which can be derived using the D1-D5 CFT, does not seem to have any ambiguity. And since we know that the dif\/ference between dif\/ferent answers are exactly accounted for by the two-centered solutions in supergravity, what can we say about the states corresponding to these solutions from the Borcherds--Kac--Moody algebra point of view?

First we begin with the f\/irst question. Although the formula (\ref{DVVPF_2}) might look unambi\-guous, the ambiguity really lies in how we expand the right-hand side of the equation. For example~\cite{0609109,Dabholkar:2007vk}, the two possible ways of expanding the following factor of the partition function
\begin{gather*}
\frac{1}{(y^{1/2}-y^{-1/2})^2}  =  \frac{1}{y(1-y^{-1})^2} = y^{-1} + 2y^{-2} + \cdots  \\
\phantom{\frac{1}{(y^{1/2}-y^{-1/2})^2}}{} =  \frac{1}{y^{-1}(1-y^{1})^2} = y^{1} + 2y^{2} + \cdots ,
\end{gather*}
corresponding to two possible ranges for the parameter $y>1$ and $y<1$, will give dif\/ferent answers for the degeneracies. This is actually a familiar phenomenon in the representations of Lie algebras. The above two expansions, for example, are exactly the two possible expansions of the square of the denominator of the $su(2)$ algebra in its two Weyl chambers.

It is not hard to convince oneself that this ambiguity of choosing expansion parameters is exactly the same ambiguity as that of choosing integration contours when we invert the equation. More generally,  we should expand the product factor in the counting formula (\ref{DVVPF_2}) in powers of~$e^{-\p i (\a,\O)}$ when $(\im\,\O, \a) < 0$ and in powers of $e^{\p i (\a,\O)}$ when $({\mathrm{Im}}\,\O, \a) > 0$. In other words, the prescription for contours is equivalent to the prescription for a way to expand the generating function.

We can give the above prescription another interpretation which makes the role of the Borcherds--Kac--Moody algebra more manifest. Let's consider the Verma module $M(\L)$ of this algebra  with highest weight $\L$ and its super-character $\text{sch}\, M(\L)$. Besides the denominator (\ref{denominator_general}), the character formula also contains a numerator. Using the formal exponential introduced earlier,
the super-character reads
\begin{gather}
\text{sch}\,  M(\L) = \sum_{\m\leq \L}\,\text{sdim}( M(\L)_\m)\,e(\m)\nonumber\\
\phantom{\text{sch}\,  M(\L)}{}
=
\frac{e(-\varrho+\L)}{e(-\varrho){\prod\limits_{\a\in \D^+}\big( 1-e(-\a)\big)^{\text{mult}\, \a}}}
=\frac{e(\L)}{\prod\limits_{\a\in \D^+}\big( 1-e(-\a)\big)^{\text{mult}\, \a}} ,\label{verma_character}
\end{gather}
where $M(\L)_\m$ denotes the $\m$-weight space of the Verma module $M(\L)$, and ``$\m\leq \L$'' means that $\L - \m$ is a sum of simple roots. The ``s'' in ``sdim'' denotes the fact that we are dealing with the graded characters counting the graded degeneracies, taking the plus or minus sign depending on whether the root involved is even or odd. Indeed, recall that we have def\/ined the number ``${\rm mult}\,\a$'' to be the graded multiplicities of the root $\a$.

Comparing this character formula with the integral (\ref{DVV_integral_2}), we note that the integrand is exactly the square of sch$M(\L)$,
namely the square of the super-character of the Verma module with highest weight
\begin{gather*}
\L= \varrho + \frac{1}{2} \L_{P,Q} ,
\end{gather*}
and the contour integral has the function of picking up the zero-weight space. Hence the dyon degeneracy $(-1)^{P\cdot Q+1} D(P,Q)$ has the interpretation of enumerating the (graded) number of ways the weight $2\L$ can be written as a sum of two copies of positive roots. Equivalently, one can interpret the dyon degeneracy as the ``second-quantized multiplicity'' of the weight $2\L$, generated by a gas of freely-acting bosonic and fermionic generators given by two copies of the positive roots of the algebra.

What we just saw is that the dyon degeneracies have a nice interpretation in terms of positive roots of the algebra which seems to be free of ambiguities, so we might wonder where the contour/moduli ambiguities we have seen from the integral/supergravity viewpoint goes in this picture.  The subtlety lies in the fact that, the character formula (\ref{verma_character}) in terms of the formal exponentials $e(-\a)$  contains the same information as the integral formula in terms of functions $e^{-\p i (\O,\a)}$ of $\O$ only  if we expand all the expressions in the latter formula in powers of $e^{-\p i (\O,\a)}$. From this consideration, the prescription for the integration contour (\ref{DVV_integral_2})
means that for this interpretation to be correct, one must make the unique choice of simple roots, and thereby the choice of positive roots, in the character formula such that the moduli vector $X$ lies in the fundamental Weyl chamber of the root system.
A manifest but crucial fact to keep in mind here is that the character formula (\ref{verma_character}) for Verma modules in {\it not} invariant under a change of simple roots.

What we have concluded from the above reasoning is that, for a super-selection sector with given total charges, a dif\/ferent choice of moduli corresponds to a dif\/ferent choice of positive roots in the algebra. This is nevertheless not very convenient. Instead we will use the equivalent description of letting the highest weight of the module be moduli-dependent while keeping the simple roots f\/ixed.

No matter whether we choose to keep the highest weight f\/ixed and vary the simple roots when we vary the moduli, or to keep the simple roots f\/ixed and vary the highest weight, from our contour condition it's clear that only in the attractor region, characterised by
\[
(X,\a) (\L_{P,Q},\a) > 0 ,\qquad \text{for all}\quad\a\in \D^{\rm re},
\]
does the counting formula correspond to a Verma module of dominant highest weight. Recall that a dominant weight is a vector lying in the fundamental Weyl chamber and having integral inner product with all roots.

For a given set of total charges, a natural choice for the simple roots is therefore such that the charge vector lies in of the fundamental domain\footnote{We ignore the special cases where the charge vector $\L_{P,Q}$ lies on the boundary of some fundamental domain, corresponding to the situation of having two-centered scaling solutions.}
\[
\L_{P,Q} \in{\cal W}.
\]
This justif\/ies our choice of conventions in the previous sections to let the charge vector $\L_{P,Q}$ lies in the fundamental Weyl chamber.

When the moduli do not lie in the attractor region, the corresponding Verma module will not have a dominant highest weight.
Indeed, when considering a point in the moduli space corresponding to
another Weyl chamber
\[
X \in w({\cal W}),\qquad w\in W,
\]
either from the contour integral (\ref{DVV_integral_2}) or from the character formula (\ref{verma_character}) we see that the dyon degeneracy is encoded in the Verma module of the following highest weight
\[
\L_w =  \varrho + w^{-1}(\L-\varrho). 
\]
Notice that the ambiguity of expansion we discuss above does not involve the imaginary positive roots, def\/ined as those positive roots that are either timelike or lightlike, which give the majority of factors in the product formula, and are therefore responsible for the asymptotic growth of the degeneracies. This is because, our $X$ is in the future lightcone by construction, the convergence criterion
\[
(X,\b) < 0 ,\qquad \b \in \D_+^{\rm im}
\]
is guaranteed to be met, independent of the precise location of $X$ inside the future light-cone. This also justif\/ies the fact that the Weyl group of a Borcherds--Kac--Moody algebra is the ref\/lection group with respect to the real roots only.

To sum up, we arrive at the conclusion that crossing a wall of marginal stability corresponds to a change of representation of the Borcherds--Kac--Moody algebra microscopically. More precisely, the change is such that the highest weight of the Verma module is related by a Weyl ref\/lection. In particular, away from the attractor region, the highest weight of the representation will no longer be dominant. If we consider the even and odd positive roots as corresponding to the bosonic and fermionic creation operators, this is described by a change of vacuum of the system.

\subsection{Microscopic derivation of the wall-crossing formula}\label{sec5.3}

In the last subsection the dyon-counting formula together with the prescription for its contour of integration leads us to
a moduli-dependent prescription of a highest weight, whose Verma module encodes the degeneracies of BPS dyons at the given moduli. In this subsection we will see what the group structure of wall-crossing implies for this microscopic description of BPS states. In particular, we will see how the wall-crossing formula predicted using the supergravity analysis can be derived from this microscopic point of view.

To begin let us consider the sequence of ref\/lections (\ref{bruhat_decay}), (\ref{bruhat_decay_2}) as described in Section~\ref{sec4.2}.

From the above discussion about the correspondence between the moduli and the highest weights appearing in the dyon-counting formula, we see that the above weak Bruhat ordering gives the following sequence of the highest weights:
\begin{gather*}
\L =\L_{w_0}\to \L_{w_1}\to \cdots \to  \L_{w_n}  .
\end{gather*}
From the fact that the charge vector $\L_{P,Q}$ lies in the root lattice and the fundamental chamber in the future light-cone, it is not dif\/f\/icult to show that
$ w^{-1}_{m}(\L_{P,Q})> w^{-1}_{m-1}(\L_{P,Q}) $
 at all steps, namely that the dif\/ference of the two and therefore the dif\/ference $\L_{w_m} -\L_{w_{m-1}} $
 is always some sum of real simple roots.

To see this explicitly,
consider two points in the moduli space inside the two neighbouring Weyl chambers $w_{m-1}({\cal W})$ and $w_{m}({\cal W})$ respectively, separated by the wall $(X,\a)=0$  with
\[
\a= w_{m-1}(\a_{i}),
\]
where $\a_i$ can be any of the three simple roots. In other words, we have
\[
w_m^{-1} = s_{i} w_{m-1}^{-1} .
\]

It will be convenient to use the root $\a$ and the corresponding light-like vectors $\a^\pm$ as a basis and
decompose the charge vector $\L_{P,Q}$ in the way introduced in (\ref{charge_split_new}).

This gives its image under the Weyl group element $w_{m-1}^{-1}$ as
\begin{gather*}
w^{-1}_{m-1} (\L_{P,Q})= P_\a^2 \a_{i}^+ + Q_\a^2 \a_{i}^- - |(P\cdot Q)_\a|  \a_i  .
\end{gather*}

By construction we have $(\a_{\pm},\a ) = 0$, which implies that the light-like vectors $\a_i^\pm$ are invariant under the ref\/lection $s_i$ and therefore
\begin{gather*}
w^{-1}_{m}(\L_{P,Q}) = P_\a^2 \a_{i}^++ Q_\a^2 \a_{i}^- + |(P\cdot Q)_\a|  \a_i  .
\end{gather*}

We have therefore shown that
\[
\L_{w_{m}} - \L_{w_{m-1}}  =\,(\L_{w_{m}}-\varrho,\a_i) \a_i.
\]

Therefore, the hierarchy (the weak Bruhat ordering) of the group elements, which have been identif\/ied with the direction given by the attractor f\/lows, induces the following hierarchy of Verma modules of the Borcherds--Kac--Moody algebra
\begin{gather}\label{sequence_module}
M(\L)= M(\L_{w_0})\subset M(\L_{w_1})\subset \cdots \subset M(\L_{w_n}) .
\end{gather}

In this sequence, every module on the left in a sub-module on the right. Explicitly, the highest weight vector of the $(m-1)$-th module is the highest weight vector of the $m$-th module acted on by the element
\[
f_{i_m}^{ (\L_{w_{m}}-\varrho, \a_{i_m}) }
\]
of the algebra. Indeed, one can show that the combination $e_j f_{i_m}^{ (\L_{w_{m-1}}-\varrho, \a_{i_m})  }
$  annihilates the highest weight vector for all $j$ and this guarantees that the next module $M(\L_{w_{m-1}})$ along the attractor f\/low is again a highest weight module.

Next we are interested in the following question: what does this hierarchy of representations of the algebra imply for the counting of BPS states when the background moduli are changed along the attractor f\/low?

Let's again concentrate on the neighbouring regions $w_{m-1}({\cal W})$ and $w_{m}({\cal W})$. We have just seen that the dif\/ference between the corresponding highest weights of the relevant Verma modules is a multiple of the real simple roots $\a_i$. For counting the dif\/ference between the number of dyonic states we want to exploit the fact that all but one factor in the product $\prod_{\b\in \D_+} (1- e(-\b))^{\text{mult}\,\b}$ is invariant under the Weyl ref\/lection $s_i$,
which can be either concluded by the fact that the Weyl ref\/lection $s_i$ ref\/lects the positive root $\a_i$ and permutes all the others (Theorem~\ref{length_posi_negi}), or by inspection of the explicit expression for the present algebra~(\ref{DVVPF_2}). This dif\/ference in  transformation properties of the positive roots under the Weyl ref\/lection with respect to the simple root $\a_i$ naturally divides the product formula into two factors
\[
\prod_{\a\in \D_+} \left(\frac{1}{1-e(-\a)}\right)^{2{\text {mult}}\,\a} =  \frac{1}{(1-e(-\a_i))^2} \times\prod_{\substack{\b\in \D_+\\ \b\neq \a_i}} \left(\frac{1}{ 1- e(-\b)}\right)^{c(|\b|^2)}.
\]

We will thus from now on use the simple root $\a_i$ and the corresponding light-like vectors $\a_i^\pm$ as our basis for the space $\R^{2,1}$. In particular, it's easy to check that the Weyl vector can be written in this basis as
\[
\varrho = \a_{i}^+ + \a_{i}^- -\frac{1}{2} \a_i  .
\]

Now consider the highest weight
\[
2 \L_{w_{m}}
= (P_\a^2 +2 )\a_i^+ + (Q_\a^2 +2 )\a_i^- + \big( |(P\cdot Q)_\a| -1\big)\a_i
\]
as the sum of two vectors $n \a_i$ and $2 \L_{w_{m}} -n\a_i$, where the f\/irst vector comes from the f\/irst factor of the product formula, corresponding to the generator $\a_i$. We can then write the dyon degeneracy at moduli space $X\in w_{m}({\cal W})$ as
\begin{gather*}
(-1)^{P\cdot Q+1}D(P,Q)\lvert_{w_{m}}
= \sum_{ n\geq0} (n+1) f(n) ,
\end{gather*}
where we have def\/ined the integer $f(n)$ to be the degeneracy of the state corresponding to the weight $2 \L_{w_{m}} -n\a_i$, generated by
generators given by all positive roots but with $\a_i$ excluded. The pre-factor is simply the degeneracy of a state of oscillation level $n$ when there are two identical harmonic oscillators. From the same consideration we have for the other side of the wall
\begin{gather*}
(-1)^{P\cdot Q+1}D(P,Q)\lvert_{w_{m-1}}
= \sum_{ n\geq0} (n+1)\,f(n+2 |(P\cdot Q)_\a|) .
\end{gather*}

Now, the fact that the Weyl ref\/lection with respect to the real simple root $\a_i$ only takes the simple root to minus itself while permuting the rest of the positive roots simply means the property of the degeneracy $f(n)$ that the two integers $n$ and $2|(P\cdot Q)_\a|-2-n$ have the same image under the map $f:\Z \to \Z$. Furthermore, the limiting expression (\ref{pole_look_like}) at the pole $(\O,\a_i)\to 0$ of the partition function is equivalent to the statement that
\begin{gather*}
\sum_{n=-\inf}^{\inf} f(n)  = d(P_\a)  d(Q_\a) .
\end{gather*}

Using these two properties of the degeneracy $f(n)$ and after some manipulation of the formula, we obtain
\begin{gather*}
D(P,Q)\lvert_{w_{m}}
=D(P,Q)\lvert_{w_{m-1}} + (-1)^{(P\cdot Q)_\a+1}  |(P\cdot Q)_\a| d(P_\a)  d(Q_\a) ,
\end{gather*}
where we have also used  $(P\cdot Q)_\a = P\cdot Q$ mod 2, which can be shown by the fact that the charge lattice $\G^{6,22}$ is even. Here we see that what we derived above from considering the dif\/ference of representations corresponding to dif\/ferent parts of the moduli space, is exactly the wall-crossing formula derived using the low-energy supergravity theory. In other words, the dif\/ference between the microscopic degeneracies in two neighbouring Weyl chambers separated by the wall $(X,\a)=0$
 is precisely the entropy carried by the two-centered supergravity solution given by~$\a$. Therefore, the wall-crossing formula which follows from our proposal for counting dyon as the second-quantized multiplicity of a moduli-dependent highest weight coincides with the two-centered wall-crossing formula derived using supergravity analysis, and in fact can be
thought of  providing a microscopic derivation of the two-centered wall-crossing formula.

\section{Conclusions and discussions}\label{sec6}

In this paper we have discussed the role of a Borcherds--Kac--Moody superalgebra in the microscopic dyon degeneracy of the $\cn=4$, $d=4$ string theory obtained by compactifying heterotic string on $T^6$, or equivalently the type II string on $K3\times T^2$. In particular we have clarif\/ied the relation between this algebra and the wall-crossing of the two-centered bound states of the theory.

First we have shown that the Weyl group of the algebra has the physical interpretation as the group of wall-crossing, or the group of discrete attractor f\/low of the theory. This is done by relying on the analysis of the low-energy ef\/fective supergravity theory which shows how the walls of the Weyl chambers can be identif\/ied with the physical walls of marginal stability in the moduli space. Especially, the ordering of the group elements induces an ordering of the two-centered solutions which gives the structure of an RG f\/low.

Second we propose that the moduli-dependent dyon degeneracy is given by enumerating the ways a charge- and moduli-dependent highest weight can be generated by a set of freely acting bosonic and fermionic generators, which are given by two copies of positive roots of the Borcherds--Kac--Moody superalgebra. In other words we propose an equivalence between the dyon degeneracy and the second-quantized multiplicity of a moduli-dependent highest weight in the algebra.
 In particular, using our identif\/ication of the Weyl group as the discrete attractor group, we show that the attractor f\/low generates a sequence of Verma sub-modules (\ref{sequence_module}), which terminates at the smallest module when all moduli have f\/lown to their attractor values. By comparing the neighbouring modules and using the properties of the multiplicities of the roots, one can furthermore derive a wall-crossing formula from our counting proposal, which exactly reproduces the wall-crossing formula obtained from supergravity consideration.

These results strongly suggest the following picture: the BPS states are generated by the generators corresponding to the positive roots of the algebra; crossing a wall of marginal stability corresponds to a change of vacuum state, or a change of representation of the algebra, or alternatively and equivalently, exchanging some pairs of creation and annihilation operators. We believe that this sheds light on both the role of the Borcherds--Kac--Moody algebra in the BPS sector of the theory and the nature of wall-crossing in this theory.
What we have not answered in the following question: can we have an interpretation of this Borcherds--Kac--Moody superalgebra from the low-energy ef\/fective action point of view or microscopic point of view? In particular, the positive real roots of the algebra have clear supergravitational interpretation as corresponding to the stationary solutions with two centers carrying $1/2$-BPS charges, but how about the positive imaginary roots which are responsible for the bulk of the black hole entropy? Alternatively, we might also ponder about the microscopic interpretation of the present algebra. While the $1/2$-BPS degeneracies have an interpretation of counting curves in the $K3$ manifold~\cite{Yau:1995mv}, our result suggests that any such interpretation for $1/4$-BPS should have an underlying Borcherds--Kac--Moody algebra.
Similarly, one expects an interpretation of this algebra in terms of the three string junctions as discussed in \cite{Gaiotto:2005hc,Sen:2007ri}.
Finally, on the gauge theory side, a similar Weyl-group-like structure also appears in Seiberg--Witten theory and in Seiberg-like dualities~\cite{Cachazo:2001sg}. It would be very interesting if any connection can be found between the appearances of the Weyl group in theories with and without gravity.

There are numerous immediate generalizations one might explore. The f\/irst one is how the algebra and the Weyl group are modif\/ied when the charges have other $T$-duality invariants being non-trivial, namely when the condition (\ref{co_prime_condition}) is no longer satisf\/ied. In this case there are extra two-centered solutions and the wall-crossing formula is modif\/ied. The second one is the generalization to other $\cn=4$ theories, in particular the CHL models \cite{w_in_progress}. Finally, the most challenging and tantalizing idea to explore is how much of the group and algebraic structure unveiled in the present paper can survive in the theories with less supersymmetries. In particular whether there is any concrete relation between the general wall-crossing formulas~\cite{Kontsevich_Soibelman} and Borcherds--Kac--Moody superalgebras.

\appendix
\section{Properties of Coxeter groups}
\label{Appendix: Properties of Coxeter Groups}

In this appendix we collect various def\/initions and facts about Coxeter groups. The proofs of them can be found in \cite{coxeter_book1,coxeter_book2,coxeter_book3}. Our presentation is similar to that of \cite{Henneaux:2007ej}.
\begin{definition}[Coxeter system]
\label{def_coxeter_grp}
A Coxeter system $(W,{\cal S})$ consists of a Coxeter group $W$ and a set of generators ${\cal S}=\{s_i, \ i = 1,\dots,n\}$, subjected to the relations
\begin{gather} \label{def_coxeter}
s_i^2  = 1 ,\qquad (s_is_j)^{m_{ij}}=1
\end{gather}
with
\begin{gather*}
m_{ij}  = m_{ji} \geq 2 \qquad \text{for} \ \ i\neq j.
\end{gather*}

A Coxeter graph has $n$ dots connected by single lines if $m_{ij} > 2$, with $m_{ij}$ written on the lines if $m_{ij} > 3$.
\end{definition}

\begin{definition}[Length function]
Def\/ine the length function
\begin{gather}\label{length_function}
\ell: W \rightarrow \Z_+
\end{gather}
such that an element has length $\ell$ if there is no way to write the element in terms of a product of less than $\ell$ generators. For $w\in G$, from
\begin{gather*}
\ell(ws_i)  \leq  \ell(w) + 1,\\
\ell(w) = \ell(ws_i^2 )   \leq  \ell(ws_i) + 1
\end{gather*}
we see that
\[
\ell(w) - 1 \leq \ell(ws_i) \leq \ell(w) + 1 .
\]

Furthermore, the length function def\/ines a distance function on the group $d: W\times W \rightarrow \Z_+$ as
\[
d(w,w') = \ell(w^{-1}w') = d(w',w) .
\]
One can easily check that this is a metric, especially that the triangle inequality is satisf\/ied.
\end{definition}

\begin{definition}[Roots of the Coxeter group]
Def\/ine the set of roots
\begin{gather}\label{root1}
\D^{\rm re} = \bigl\{ w(\a_i) , \ w\in W,\  i = \{1,\dots,n\} \bigr\}
\end{gather}
especially $\a_i$'s are called the {\it simple roots}\footnote{The notation $\D^{\rm re}$ is adapted to the fact that the real roots of the set of Borcherds--Kac--Moody algebra is the set of roots of the Weyl Coxeter group.}. A root
\[
\a = \sum_i a^{(i)} \a_i
\]
is called a {\it positive root} if all $a^{(i)}\geq0$ and a {\it negative root} if all $a^{(i)}\leq0$. We will denote them as $\a>$ and $\a<$ respectively.
\end{definition}

\begin{theorem}\label{length_posi_negi}
\[
\ell(ws_i) = \ell(w) +1
\]
iff $w(\a_i)$ is a positive root.
\end{theorem}

\begin{corollary}
A root is either positive or negative.
\end{corollary}

\begin{theorem}
For hyperbolic Coxeter group, the Tits cone, namely the image of a connected fundamental domain under the group action, is the future light-cone. Projected onto a constant length surface, this gives a tessellation of the hyperbolic space.
\end{theorem}

\begin{definition}[Bruhat order]
\label{bruhat_order}
Given the Coxeter system ($W,S$) and its set of ref\/lections
${\cal R} = \{w s_i w^{-1}\lvert\, w\in W ,\  s_i \in {\cal S}\}$, and let $u,u' \in W$, then
\begin{enumerate}\itemsep=0pt
\item $u\rightarrow u'$ means that $u^{-1}u' \in {\cal R}$ and $\ell(u)<\ell(u')$;
\item $u \leq_B u''$ means that there exists $u_k \in W$ such that
\begin{gather*}
u \rightarrow u_1 \rightarrow \cdots \rightarrow u_m \rightarrow u''.
\end{gather*}
\end{enumerate}
\end{definition}

\begin{definition}[Weak Bruhat order]
Given two elements $u,u' \in W$, repeating the above def\/inition but now restrict further to
\begin{gather}
 \label{weak_order} u^{-1}u' \in {\cal S},
 \end{gather}
we obtain the ``weak Bruhat order'' $u\leq_{wB} u''$.
\end{definition}

\subsection*{Acknowledgments}

We would like to thank Atish Dabholkar, Frederik Denef, Axel Kleinschmidt, Greg Moore, Daniel Persson, Boris Pioline and Curum Vafa for useful discussions.
E.V. would like to thank Harvard University for hospitality during the
completion of this work. M.C. is supported by the Netherlands Organisation for Scientif\/ic Research (NWO).
The research of E.V. is partly supported by the Foundation of Fundamental Research on Matter (FOM).

\pdfbookmark[1]{References}{ref}
\LastPageEnding

\end{document}